\documentclass[%
 reprint,
 amsmath,amssymb,
 aps,
]{revtex4-2}

\usepackage{todonotes}
\usepackage{graphicx}
\usepackage{dcolumn}
\usepackage{subfigure}
\usepackage[normalem]{ulem}
\usepackage{bm}

\usepackage{todonotes}

\begin{document}

\preprint{APS/123-QED}

\title{Unsteady dynamics of a classical particle-wave entity}

\author{Rahil N. Valani$^{1}$}\email{rahil.valani@gmail.com}
\author{Anja C. Slim$^{2,3}$}
\author{David M. Paganin$^{1}$}%
\author{Tapio P. Simula$^{4}$}
\author{Theodore Vo$^{2}$}

\affiliation{$^1$School of Physics and Astronomy, Monash University, Victoria 3800, Australia} 
\affiliation{$^2$School of Mathematics, Monash University, Victoria 3800, Australia}
\affiliation{\mbox{$^3$School of Earth, Atmosphere and Environment, Monash University, Victoria 3800, Australia}}
\affiliation{$^4$Optical Sciences Centre, Swinburne University of Technology, Melbourne 3122, Australia}


\date{\today}

\begin{abstract}

A droplet bouncing on the surface of a vertically vibrating liquid bath can walk horizontally, guided by the waves it generates on each impact. This results in a self-propelled classical particle-wave entity. By using a one-dimensional theoretical pilot-wave model with a generalized wave form, we investigate the dynamics of this particle-wave entity. We employ different spatial wave forms to understand the role played by both wave oscillations and spatial wave decay in the walking dynamics. We observe steady walking motion as well as unsteady motions such as oscillating walking, self-trapped oscillations and irregular walking. We explore the dynamical and statistical aspects of irregular walking and show an equivalence between the droplet dynamics and the Lorenz system, as well as making connections with the Langevin equation and deterministic diffusion.
\end{abstract}

\maketitle


\section{Introduction}
Vertically vibrating a bath of liquid can result in the emergence of a self-propelled particle-wave entity in the form of a walking droplet on the free surface of the liquid~\citep{Couder2005,Couder2005WalkingDroplets,Molacek2013DropsTheory,superwalker}. The walking droplet, also known as a walker, on each bounce locally generates a slowly decaying standing wave. The droplet then interacts with these self-generated waves on subsequent bounces to propel itself horizontally. The walker emerges for vibration amplitudes just below the Faraday instability threshold where the liquid surface remains flat everywhere except in the vicinity of the walker; above this threshold the whole interface becomes unstable to standing Faraday waves~\citep{Faraday1831a}. Very close to but below the Faraday threshold, the waves created by a walker on each bounce extend far in space and decay very slowly in time. In this regime, the droplet is not only influenced by the wave it created from its most recent bounce, but also by the waves it created in the distant past, giving rise to \emph{memory} in the system. 

In the high-memory regime, walkers have been shown to mimic several peculiar features that were previously thought to be exclusive to the quantum realm. These include orbital quantization in rotating frames \citep{Fort17515,harris_bush_2014,Oza2014} and harmonic potentials \citep{Perrard2014b,Perrard2014a,labousse2016}, Zeeman splitting in rotating frames \citep{Zeeman,spinstates2018}, wavelike statistical behavior in confined geometries \citep{PhysRevE.88.011001,Giletconfined2016,Saenz2017,Cristea,durey_milewski_wang_2020} as well as in an open system \citep{Friedal} and tunneling across submerged barriers \citep{Eddi2009,tunnelingnachbin,tunneling2020}. Walkers have also been predicted to show anomalous two-droplet correlations \citep{ValaniHOM,correlationnachbin}. Recently, efforts have also been made to develop a hydrodynamic quantum field theory for the walking-droplet system~\citep{Dagan2020hqft,Durey2020hqft}. Detailed reviews of hydrodynamic quantum analogs of walking droplets have been provided by \citet{Bush2015} and \citet{Bush_2020}. 

To model the walking droplet, many theoretical descriptions have been developed over the years. These range from phenomenological stroboscopic models that average over the vertical periodic bouncing motion of the droplet and only capture the horizontal dynamics, to sophisticated models that resolve the vertical and horizontal dynamics and the detailed evolution of the surface waves created by the walker. A review of the different models is given by \citet{Turton2018} and \citet{Rahman2020review}. The latter work provides a perspective through the lens of dynamical systems theory. 

In experiments, a single walker or superwalker~\citep{superwalker,superwalkernumerical} is typically observed to travel in a straight line at a constant speed unless it encounters obstacles or other droplets. At high memories, \citet{Bacot2019} experimentally observed multiple states of a free walker, where in addition to rectilinear constant speed motion, the droplet was also observed to walk with oscillations in speed in the walking direction. Using a theoretical model of walkers, \citet{Hubert2019} showed that in the very-high-memory regime, the rectilinear constant speed motion of a walker becomes unstable and the walker's horizontal dynamics becomes bimodal where it erratically switches between phases of linear motion and diffusive motion~\citep{Hubert2019}. This bimodal motion shows analogies with the run-and-tumble dynamics common in swimming micro-organisms and artificial microswimmers~\citep{BERG1972,stopgoswim,Stocker2635,Bhattacharjee2019}.

\citet{Durey2020} also explored this high-memory regime for a walker using the stroboscopic model of \citet{Oza2013}, by confining the motion of the walker to a line. Since the steady walking is neutrally stable to lateral perturbations~\citep{Oza2013}, the key aspects of the instability of the steady walking state may be captured by investigating the droplet's dynamics confined to one dimension~\citep{phdthesisrahil,Durey2020}. They identified various regimes of a walker in the parameter space, that give rise to oscillations in the walking speed and random-walk-like motion of the droplet, leading to a statistical wavelike signature in the probability density function of the droplet's position. Wavelike statistics emerging from speed oscillations of a walker has also been demonstrated in a hydrodynamic analog of Friedel oscillatio  ns~\citep{Friedal}. {Recently, \citet{Durey2020lorenz} investigated the high-memory regime using an idealized theoretical pilot-wave model that implements a simplified sinusoidal wave form for the waves generated by the droplet and found similarities between the droplet's dynamical system and the Lorenz system.}

In this paper, we revisit the dynamics of a single walker restricted to move in one horizontal dimension, by extending the stroboscopic model of \citet{Oza2013} to a generalized wave form, i.e. allowing a general spatial structure for the underlying waves generated by the droplet, and exploring the dynamics observed in the parameter space using different wave forms. We investigate the role played by spatial decay of the wave form and wave oscillations in the droplet's dynamics by employing a Gaussian, Bessel and a sinusoidal wave form, and explore in detail the unsteady dynamics arising from a sinusoidal wave form. In Sec.~\ref{model} we present the generalized stroboscopic model and use it to perform a linear stability analysis for both the stationary state and the steady walking state of the droplet, in Secs.~\ref{stationary} and \ref{Walk} respectively. We then in Sec.~\ref{PS} explore the various unsteady behaviors observed in the parameter space using different wave forms. In Secs.~\ref{Dynamics irr} and \ref{Statistics irr} we explore, respectively, the dynamical and the statistical aspects of the irregular walking motion realized in the unsteady walking regime and draw connections with the Lorenz equations, the Langevin equation and deterministic diffusion.  

\section{Stroboscopic model with a generalized wave form}\label{model}

\begin{figure}
\centering
\includegraphics[width=\columnwidth]{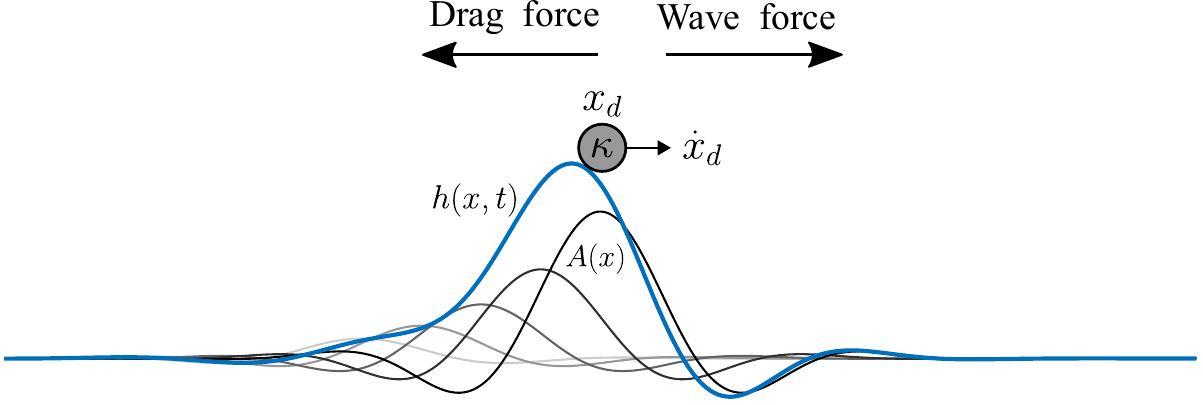}
\caption{Schematic of the walking droplet particle-wave system, showing a droplet of dimensionless mass $\kappa$ located at $x_d$ and walking horizontally with velocity $\dot{x}_d$. The droplet experiences a wave force, $-\beta\,\partial h/\partial x|_{x=x_d}$, from the underlying wave field $h(x,t)$ (blue curve) and a drag force, $-\dot{x}_d$. The underlying wave field $h(x,t)$ is the superposition of the individual waves (black and gray curves), of the spatial form $A(x)$ and decaying exponentially in time, that are continuously generated by the droplet along its trajectory.}
\label{Fig: scheme}
\end{figure}

\citet{Oza2013} derived a stroboscopic model to describe the horizontal dynamics of a walking droplet by averaging over its vertical periodic bouncing motion and employing a Bessel function of the first kind and zeroth order, $\text{J}_0(\cdot)$, wave form for the individual waves generated by the droplet on each bounce. {Here we extend this model to an arbitrary smooth, symmetric standing wave form with exponential temporal decay, and investigate the droplet's dynamics by restricting the horizontal motion of the droplet to one dimension.}

As shown schematically in Fig.~\ref{Fig: scheme}, consider a droplet at position $x_d$ walking horizontally with velocity $\dot{x}_d$ and continuously generating waves with prescribed spatial structure $A(x)$ that decay exponentially in time. The equation of motion governing the horizontal dynamics of the droplet is given by, 
\begin{equation}\label{eq: traj_1}
\kappa\ddot{x}_d+\dot{x}_d
=-\beta\frac{\partial h}{\partial x}\Big{|}_{x=x_d}.
\end{equation}
The left hand side of this equation comprises an inertial term $\kappa\ddot{x}_d$ and a drag term $\dot{x}_d$, where the overdot denotes differentiation with respect to time $t$. The right hand side of the equation captures the forcing on the droplet by the underlying wave field $h(x,t)$. This force is proportional to the gradient of the underlying wave field. The shape of the wave field $h(x,t)$ is calculated through integration of the individual wave forms $A(x)$ that are continuously generated by the particle along its trajectory. This gives 
\begin{equation}\label{eq: traj_2}
h(x,t)=\int_{-\infty}^{t}A(x - x_d(s))\,\text{e}^{-(t-s)}\,\text{d}s.
\end{equation}
Combining Eqs.~\eqref{eq: traj_1} and \eqref{eq: traj_2} we obtain the integro-differential equation,
\begin{align}\label{eq_1}
\kappa\ddot{x}_d+\dot{x}_d
=\beta\int_{-\infty}^{t}f(x_d(t) - x_d(s))\,\text{e}^{-(t-s)}\,\text{d}s,
\end{align}
where $f(x)=-A'(x)$ is the negative gradient of the wave form and the prime denotes differentiation with respect to the argument $x$. {The two parameters in this dimensionless equation of motion, $\kappa >0$ and $\beta >0$, follow directly from \citet{Oza2013} and may be usefully interpreted as the ratio of inertia to drag and the ratio of wave forcing to drag respectively.  We note that $\kappa \sim 1/\text{Me}$ and $\beta \sim \text{Me}^2$, where $\text{Me}$ is the memory parameter which represents the proximity to the Faraday threshold~\citep{Oza2013}.}

\section{The stationary solution and its linear stability analysis}\label{stationary}

We start by seeking stationary solutions of~Eq.~\eqref{eq_1}. Substituting $x_d(t)=x_0$ in Eq.~\eqref{eq_1} we arrive at the condition
\begin{equation*}
    f(0)=-A'(0)=0.
\end{equation*}
Since the wave form $A(x)$ is assumed to be smooth and symmetric, the above equation is always satisfied and we have a stationary solution. 

To determine the stability of the stationary solution, we follow the linear stability approach taken in \citet{Oza2013} and apply a perturbation $x_d(t)=x_0+\epsilon x_1(t) H(t)$ to the stationary solution. Here $H(\cdot)$ is the Heaviside step function introduced to apply the perturbation for $t\geq 0$, and {$\epsilon>0$ is a small perturbation parameter}. Substituting this in Eq.~\eqref{eq_1}, we find that the perturbation, $x_1$, evolves according to
\begin{align*}
    \kappa\ddot{x}_1+\dot{x}_1
&=\beta f'(0) \Big[x_1(t)-\int_{0}^{\infty}x_1(t-z)H(t-z)\,\text{e}^{-z}\,\text{d}z\Big].
\end{align*}

Taking the Laplace transform of both sides results in
\begin{equation}\label{poles ss}
X_{1}(s)=\frac{(s+1)(\kappa (s x_{1}(0)+\dot{x}_{1}(0))+x_{1}(0))}{s[\kappa s^2 + (1+\kappa) s + 1-\beta f'(0)]}.
\end{equation}
In Eq.~\ref{poles ss}, $X_1(s)=\mathcal{L}\{x_1(t)\}$ with $\mathcal{L}\{\cdot\}$ denoting the Laplace transform from the time domain $t$ to the (complex) frequency domain $s$. The stability of the stationary state can be determined by finding the poles of $X_{1}(s)$ in Eq.~\eqref{poles ss}~\citep{Oza2013}. In the present case, the poles are the roots of
\begin{equation*}
s[\kappa s^2 + (1+\kappa) s + 1-\beta f'(0)]=0.
\end{equation*}
The trivial $s=0$ solution corresponds to translation invariance of the system. The two non-trivial poles are the solution of the quadratic factor.
The stationary state becomes unstable when one of the non-trivial poles become positive.
%
This takes place when
\begin{equation*}
    \beta=\frac{1}{f'(0)}.
\end{equation*}

\begin{figure}
\centering
\includegraphics[width=\columnwidth]{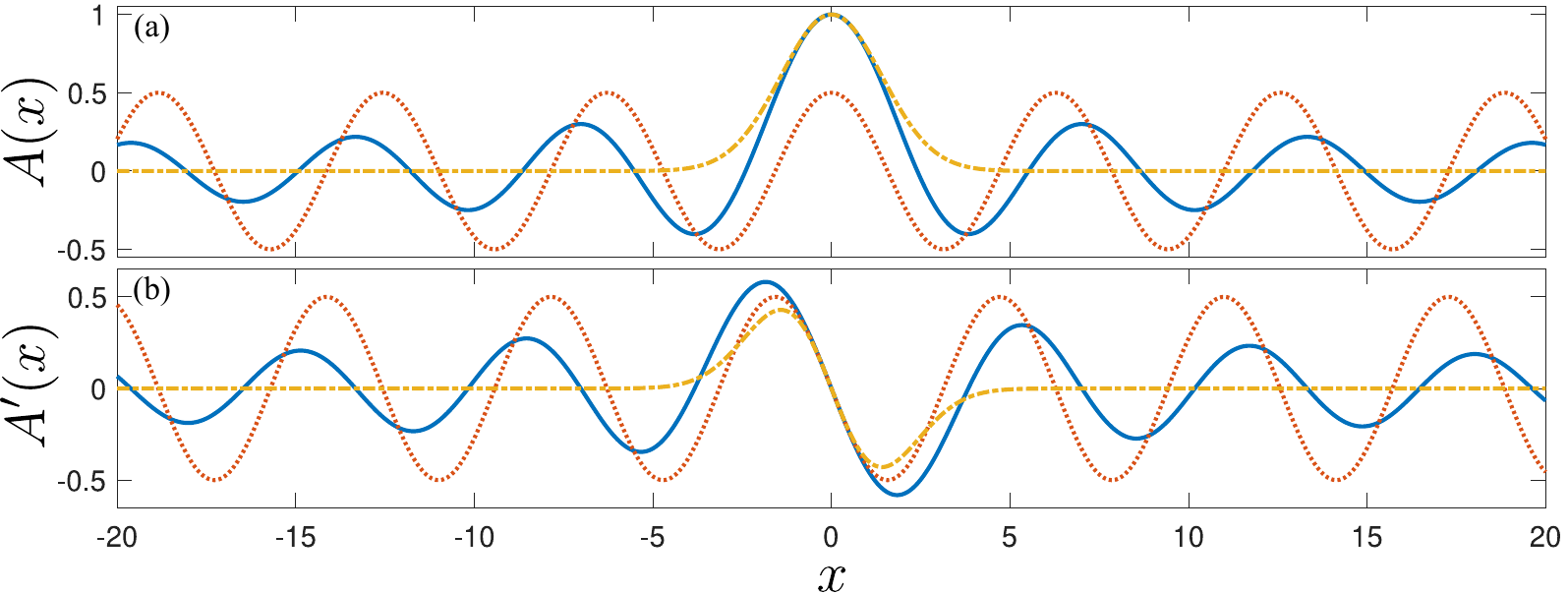}
\caption{Comparison of the following three different (a) wave forms $A(x)$ and (b) their gradients: a Gaussian wave form $\text{e}^{-(x/2)^2}$ (yellow dash-dotted curve), a Bessel function wave form $\text{J}_0(x)$ (blue solid curve), and a sinusoidal wave form $\cos(x)/2$ (red dotted curve).}
\label{Fig: wave compare}
\end{figure}

\section{The steady walking solution and its linear stability analysis}\label{Walk}

\begin{figure*}
\centering
\includegraphics[width=2\columnwidth]{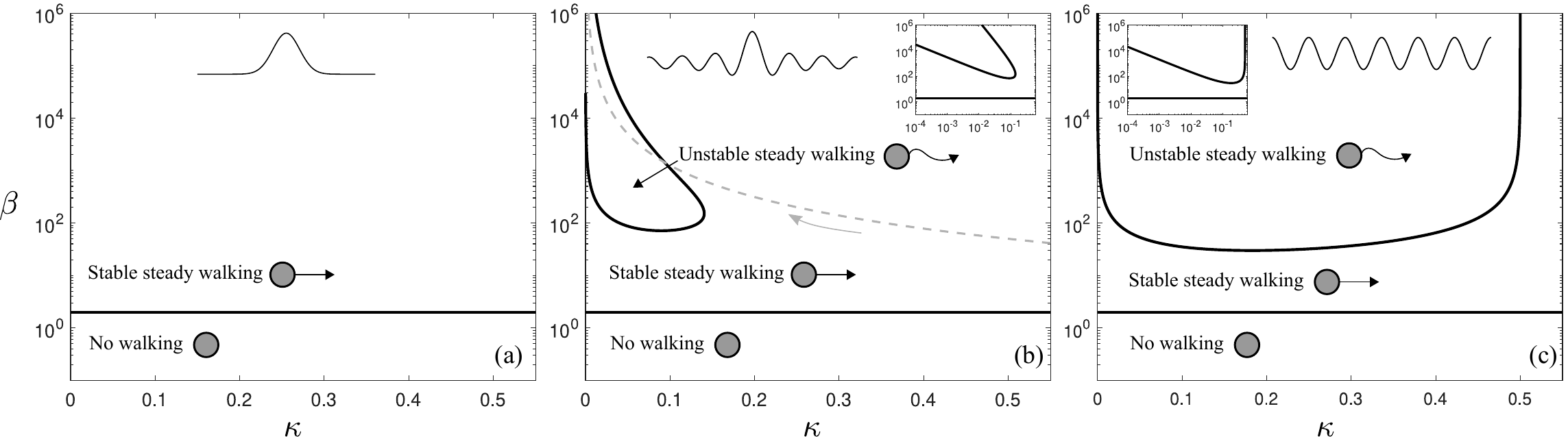}
\caption{Linear stability  in the ($\kappa,\beta$) parameter space for inline perturbations to the steady walking solution of a single droplet using (a) a Gaussian wave form $\text{e}^{-(x/2)^2}$, (b) a Bessel wave form $\text{J}_0(x)$ and (c) a sinusoidal wave form $\cos(x)/2$. The gray dashed curve in (b) shows the path traversed in the parameter space for typical experimental parameters as the driving acceleration (or the memory) is increased. In each of (b) and (c), the inset shows the instability boundary using a logarithmic scale in both horizontal and vertical directions.}
\label{fig: lin stab bessel}
\end{figure*}

Once the stationary state becomes unstable for $\beta>1/f'(0)$, we obtain a steady walking state. We look for a steady walking solution with speed $u$ of the generalized stroboscopic model by substituting $x_d(t)=ut$ in Eq.~\eqref{eq_1}, which results in
\begin{equation} \label{eq:speed_constraint}
u=\beta \int_{0}^{\infty} f(uz)\,\text{e}^{-z}\,\text{d}z.
\end{equation}
By making a change of variables $uz=r$ in the integral, we can rewrite this as
\begin{equation}\label{eq: steady walk}
u^2=\beta \int_{0}^{\infty} f(r)\,\text{e}^{-r/u}\,\text{d}r=\beta F\left(\frac{1}{u}\right),
\end{equation}
where $F(s)$ is the Laplace transform of $f(r)$. Provided that a solution to the above equation exists, we obtain the steady walking speed $u$ of the droplet for a given parameter $\beta$ and spatial wave form $A(x)$ or equivalently its gradient function $f(x)$.

To determine the stability of the steady walking solution in this generalized framework, we follow \citet{Oza2013} and apply a perturbation of the form $x_d(t)= u t + \epsilon x_{1}(t) H(t)$ to the steady walking solution with speed $u$. By substituting this in Eq.~\eqref{eq_1} and comparing the $O(\epsilon)$ terms, we get
\begin{align}\label{eq: per_gen}
\kappa \ddot{x}_1+\dot{x}_1
&=\beta \Big[ x_1(t)\int_{0}^{\infty}f'(uz)\,\text{e}^{-z}\,\text{d}z\\ \nonumber
&-\int_{0}^{\infty}f'(uz)x_1(t-z)H(t-z)\,\text{e}^{-z}\,\text{d}z\Big].
\end{align}
Integrating the first integral term on the right side by parts gives
\begin{align*}
\int_{0}^{\infty}f'(uz)\text{e}^{-z}\,\text{d}z&=-\frac{f(0)}{u}+\frac{1}{u}\int_{0}^{\infty}f(uz)\,\text{e}^{-z}\,\text{d}z=\frac{1}{\beta} , 
\end{align*}
where the constraint in Eq.~\eqref{eq:speed_constraint} has been used.
Substituting this in Eq.~\eqref{eq: per_gen} and Laplace transforming, we have
\begin{equation}\label{eq: poles}
    X_{1}(s)=\frac{\kappa [s x_{1}(0)+\dot{x}_{1}(0)]+x_{1}(0)}{\kappa s^2 + s -1+(s+1)\left[ F((s+1)/u)/F(1/u) \right]}.
\end{equation}
As before, the stability of the inline walking motion can be determined by finding the poles of $X_{1}(s)$ in Eq.~\eqref{eq: poles}~\cite{Oza2013}.

To model the walking dynamics of the droplet, a Bessel function wave form, $A(x)=\text{J}_0(x)$, is typically used and has been studied in detail~\citep{Oza2013,Durey2020}. This Bessel wave form has two key features: (i) a spatial decay and (ii) spatial oscillations. We decouple these features by considering two alternate wave forms: a Gaussian wave form $A(x)=\text{e}^{-(x/2)^2}$, which has spatial decay but no oscillations, and a sinusoidal wave form $A(x)=\cos(x)/2$, which has oscillations but no spatial decay. Both of these wave forms have been chosen such that their first and second derivatives, $f(x)$ and $f'(x)$, match with the Bessel function wave form at the location where the wave is created. A comparison of the three wave forms and their gradients is shown in Fig.~\ref{Fig: wave compare}. We investigate the linear stability of steady walking using a Gaussian wave form in Sec.~\ref{sec: gauss wave}, a Bessel wave form in Sec.~\ref{bessel wave form}, and a sinusoidal wave form in Sec.~\ref{sinusoidal wave form}. To understand the effects of spatial decay and spatial oscillations on the instability of the steady walking state, we also consider in Sec.~\ref{sine gauss} a combined sinusoidal Gaussian wave form $A(x)=\frac{1}{2}\cos(x)\text{e}^{-(x/2l)^2}$ and study the instability of the steady walking state as the spatial decay length scale $l$ is varied.

\subsection{A Gaussian wave form}\label{sec: gauss wave}

Choosing a Gaussian wave form, $A(x)=\text{e}^{-(x/2)^2}$, results in $f(x)=(x/2)\text{e}^{-(x/2)^2}$ in Eq.~\eqref{eq_1}. The corresponding equation for steady walking speed $u$ can be obtained from Eq.~\eqref{eq: steady walk} with
\begin{equation*}
    F\left(\frac{1}{u}\right)= 1-\frac{\sqrt{\pi}\text{e}^{1/u^2}\text{erfc}(1/u)}{u}.
\end{equation*}
In the limit of large $\beta$, the speed scales as $u\sim \sqrt{\beta}$.

On performing the linear stability analysis by numerically solving for the poles of $X_1(s)$ in Eq.~\eqref{eq: poles}, we find that the steady walking solution always remains stable. Hence we observe two qualitatively different behaviors when a Gaussian wave field is considered (see Fig.~\ref{fig: lin stab bessel}(a)): (i) No walking for $\beta\leq 1/f'(0)=2$ and (ii) stable steady walking for $\beta>2$. 

\subsection{A Bessel wave form}\label{bessel wave form}

Choosing a Bessel function wave form, $A(x)=\text{J}_0(x)$, results in $f(x)=-A'(x)=\text{J}_1(x)$. Hence,
\begin{equation*}
F\left(\frac{1}{u}\right)=\frac{u^2}{1+u^2+\sqrt{1+u^2}}.
\end{equation*}
Substituting this in  Eq.~\eqref{eq: steady walk} for the steady walking speed, we obtain~\citep{Oza2013}
\begin{equation*}
u=\frac{1}{\sqrt{2}}\sqrt{-1+2\beta-\sqrt{1+4\beta}}\,\,.    
\end{equation*}

For $\beta \leq 1/f'(0) = 2$, the stationary droplet solution is stable, while for $\beta>1/f'(0)=2$ the steady walking solution is realized. For $\beta\gg 1$, the above equation for the walking speed can be approximated by $u\approx\sqrt{\beta}$.

The linear stability analysis requires solving for the poles of $X_1(s)$ in Eq.~\eqref{eq: poles}. 
This results in solving the equation~\citep{Oza2013}
\begin{align}\label{eq: lin stab bessel}
&(\kappa s^2 +s -1)\sqrt{u^2+(s+1)^2}\left(s+1+\sqrt{u^2+(s+1)^2}\right)\nonumber \\
&+\beta(s+1)=0.
\end{align}

For small $\kappa$ and large $\beta$, a complex conjugate pair of poles cross the imaginary axis, i.e., pass through $\text{Re}(s)=0$, resulting in a change in the stability of the steady walking solution. We can find the stability boundary of the steady walking solution in the $(\kappa,\beta)$ parameter space by setting $\text{Re}(s)=0$ and substituting $s=i\omega$ in Eq.~\eqref{eq: lin stab bessel}. The linear stability diagram is shown in Fig.~\ref{fig: lin stab bessel}(b). We see that a lobe-shaped region appears for small $\kappa$ and large $\beta$, where steady walking is unstable. For a fixed $\kappa\lesssim 0.14$, as $\beta$ is increased, we get stable steady walking for small $\beta$, unstable steady walking for moderately large $\beta$ and recover the stable steady walking state for very large $\beta$. In typical experiments with walkers and superwalkers, as the driving acceleration amplitude (or equivalently the memory) is increased, the path traversed in the $(\kappa,\beta)$ parameter space is shown by a gray dashed curve in Fig.~\ref{fig: lin stab bessel}(b). {The linear stability boundary for the Bessel wave form has been analyzed in detail by \citet{Durey2020}, using a different dimensionless form for the droplet's equation of motion.}

\subsection{A sinusoidal wave form}\label{sinusoidal wave form}

Choosing a sinusoidal wave form, $A(x)=\cos(x)/2$, results in $f(x)=\sin(x)/2$ and we get
\begin{equation*}
    F\left(\frac{1}{u}\right)=\frac{u^2}{2(1+u^2)}.
\end{equation*}
Thus, the steady walking speed, as determined by Eq.~\eqref{eq: steady walk}, is
\begin{equation*}
u=\sqrt{\frac{\beta}{2}-1}.
\end{equation*}

The linear stability analysis requires solving for the poles of $X_1(s)$ in Eq.~\eqref{eq: poles}, which results in the equation
\begin{equation}\label{eq: lin stab sine wave}
 (\kappa s^2+s-1)(2s^2+4s+\beta)+\beta(s+1)=0.
\end{equation}

Similar to the Bessel wave form in Sec.~\ref{bessel wave form}, we can find the stability boundary of the steady walking solution by setting $\text{Re}(s)=0$ and substituting $s=i\omega$ in Eq.~\eqref{eq: lin stab sine wave}. This gives
\begin{equation}
\beta=\frac{2(1+4\kappa)}{\kappa(1-2\kappa)}
\end{equation}
as the instability boundary in the parameter space (see Fig.~\ref{fig: lin stab bessel}(c)) with an oscillation frequency of 
\begin{equation*}
 \omega^2=\frac{\beta-2}{2\kappa+1}
\end{equation*}
at the onset of instability. {A similar analysis of the steady walking solution and linear stability for the sinusoidal wave form, using a different dimensionless form of the equation of motion, was performed by \citet{Durey2020lorenz}.} 

\begin{figure}
\centering
\includegraphics[width=\columnwidth]{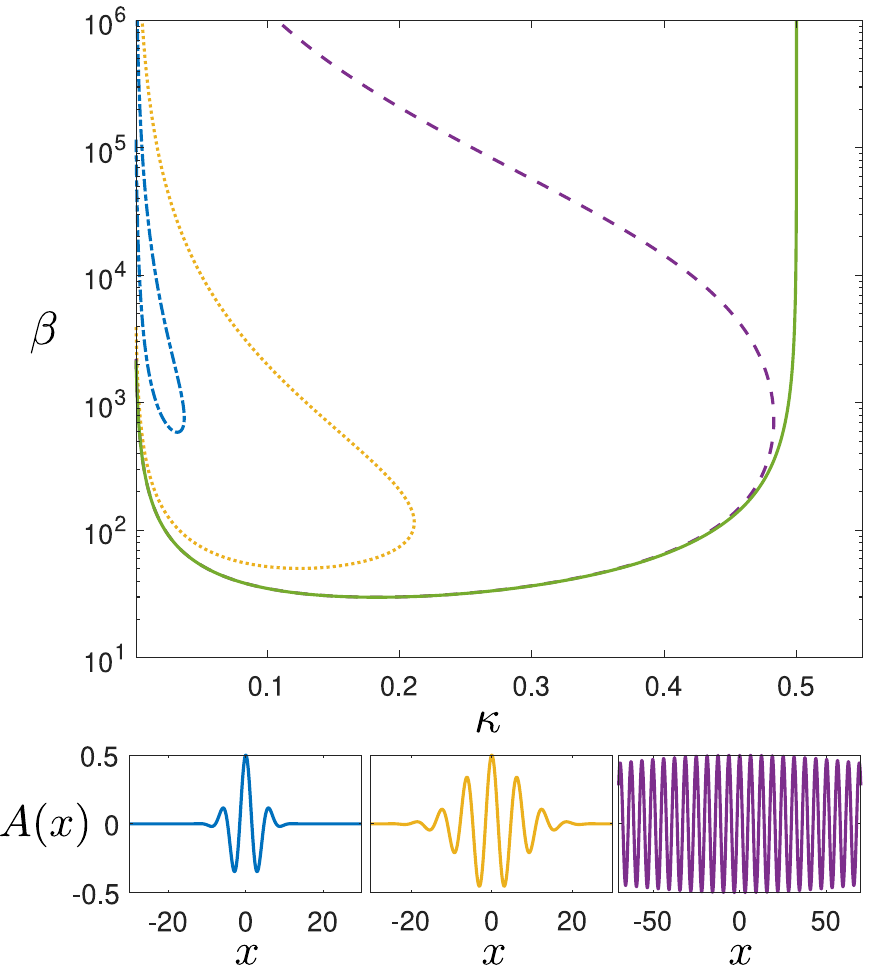}
\caption{Instability boundary of the steady walking solution in the $(\kappa,\beta)$ parameter space for a wave form $A(x)=\cos(x)\, \text{e}^{-(x/2 l)^2}\!/2$. Instability boundary (top panel) and the corresponding wave forms (bottom panel) are shown for $l=2.5$ (blue dash-dotted curve), $l=5$ (yellow dotted curve), $l=100$ (purple dashed curve), and $l \rightarrow \infty$ (green solid curve, wave form not shown).}
\label{fig: sinegauss PS}
\end{figure}

\subsection{A Sinusoidal wave form with a Gaussian envelope}\label{sine gauss}

To further understand the effect of spatial decay of the wave form on the structure of the instability boundary, we consider a wave form $A(x)=\cos(x)\,\text{e}^{\,-(x/2l)^2}\!/2$
and investigate how the instability boundary in the $(\kappa,\beta)$ parameter space is changed as the decay length scale $l$ is varied.

Figure~\ref{fig: sinegauss PS} shows the instability boundary for the steady walking state for various values of $l$ along with the wave forms. We find that for large $l$, the spatial decay is very small and it does not have a significant effect on the lower boundary of the instability. However, we do see a qualitative change in the upper boundary. For a pure sinusoidal wave form that has no spatial decay, we find that the steady walking solution is always unstable for a small range of $\kappa$ values as $\beta \rightarrow \infty$. For a wave form with a small but non-zero spatial decay, we recover stability of the steady walking state for large $\beta$. Conversely, for small $l$, we find that the region of instability shrinks rapidly with an increase in the spatial decay, indicating that oscillations in the wave form are necessary for the steady walking state to become unstable.  

Thus, by homotoping from the sinusoidal wave form to an exponentially decaying sinusoid with strong decay, we have demonstrated that the oscillations in the wave form are the key dynamic mechanism responsible for the instability of the steady walker. These oscillations play important roles in the unsteady walker regime enclosed by the lobe-shaped instability boundary (Fig.~\ref{fig: sinegauss PS}), as we will show in the next section.

\section{Unsteady walking dynamics in the $(\kappa,\beta)$ parameter-space}\label{PS}

Once the steady walking state becomes unstable in the $(\kappa,\beta)$ parameter space, a variety of unsteady motions are realized. We explore the resulting unsteady dynamics in the parameter space for the Bessel and the sinusoidal wave forms.

\subsection{Bessel wave form}

\begin{figure*}
\centering
\includegraphics[width=2 \columnwidth]{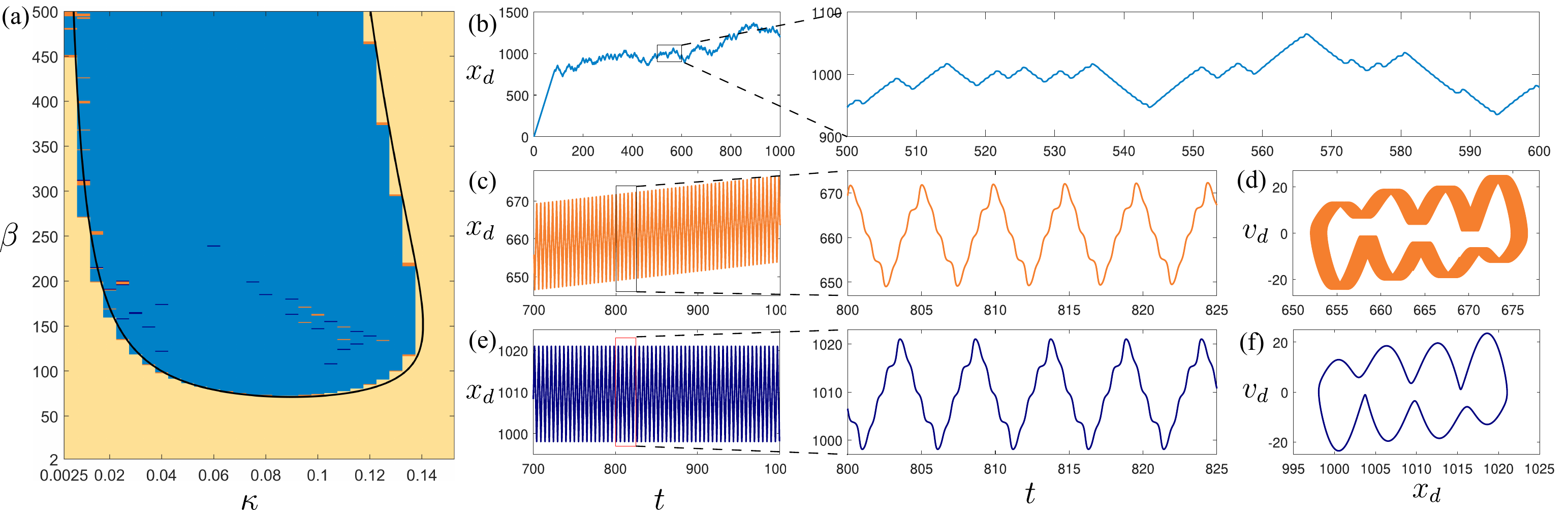}
 \caption{Walking behaviors for a Bessel wave form.  (a) Different dynamical behaviors observed in the $(\kappa,\beta)$ parameter space at $t=1000$ from simulations initiated at $t=0$ with the droplet in the steady walking state for $t\leq 0$. We explore the parameter space region $0.005\leq \kappa \leq 0.15$ and $0 < \beta \leq 500$ with resolution $\Delta \kappa=0.005$ and $\Delta \beta=1$. We observe steady walking (beige), oscillating walking (orange), self-trapped oscillations (navy blue) and irregular walking (blue). The solid black curve is the linear stability curve separating the steady walking and the unsteady walking regime. Typical trajectories of (b) irregular walking ($\kappa=0.10$, $\beta=101$), (c) oscillating walking ($\kappa=0.11$, $\beta=149$), and (e) self-trapped oscillations ($\kappa=0.12$, $\beta=139$) are shown along with the phase-space plots for (d) oscillating walking and (f) self-trapped oscillations.}
\label{fig: Ps space bessel wave}
\end{figure*}

We have explored the unsteady dynamics of a walker by numerically integrating Eq.~\eqref{eq_1} with the Bessel wave form in the $(\kappa,\beta)$ parameter space and the results are presented in Fig.~\ref{fig: Ps space bessel wave}. The simulations were initialized with the droplet in the steady walking state for $t<0$. We refer the reader to Appendix~\ref{numericsbessel} for details of the numerical implementation. 

 We identify three distinct unsteady walking regimes from simulations. These are (i) irregular walkers, (ii) oscillating walkers, and (iii) self-trapped oscillations. The first of these predominates. A typical trajectory of an irregular walker is shown in Fig.~\ref{fig: Ps space bessel wave}(b). Here, the droplet performs oscillations while walking and switches the walking direction erratically. In  small, isolated regions of the parameter space, we observe oscillating walkers and self-trapped oscillations. In the oscillating walker state, the droplet drifts while undergoing oscillations in the walking direction (see Fig.~\ref{fig: Ps space bessel wave}(c)). The oscillating walkers are reminiscent of the experimentally observed velocity oscillations of a walker at high memory by \citet{Bacot2019}. In the self-trapped oscillation state, the droplet traps itself under its self-generated wave field and performs periodic back-and-forth motion with no net drift (see Fig.~\ref{fig: Ps space bessel wave}(e)). The phase space dynamics for self-trapped oscillations shows a closed loop due to the periodic nature of the oscillations, while for oscillating walkers, we see a drift of the closed loop (see Figs.~\ref{fig: Ps space bessel wave}(f) and (d)).  \citet{Durey2020} also reported irregular and oscillating walkers (referred to as jittering modes in their paper) in their parameter space exploration of the droplet dynamics using a Bessel wave form. In addition, here we also observe stable self-trapped oscillations that were not reported previously. However, we note that similar self-trapped periodic oscillations were also observed by \citet{durey2018}, using their one-dimensional discrete-time pilot-wave model. Self-trapped states have also been observed when the walker is free to move in two horizontal dimensions. Here, the walker's self-generated wave field confines itself to a circular orbit~\citep{Fort_2013,labousse2016,Spinstates}. We note that, as shown in the supplemental material of \citet{Durey2020lorenz}, we also observe hysteresis near the boundary separating steady and unsteady walking. 
 
\subsection{Sinusoidal wave form}

\begin{figure*}[!t]
\centering
\includegraphics[width=2\columnwidth]{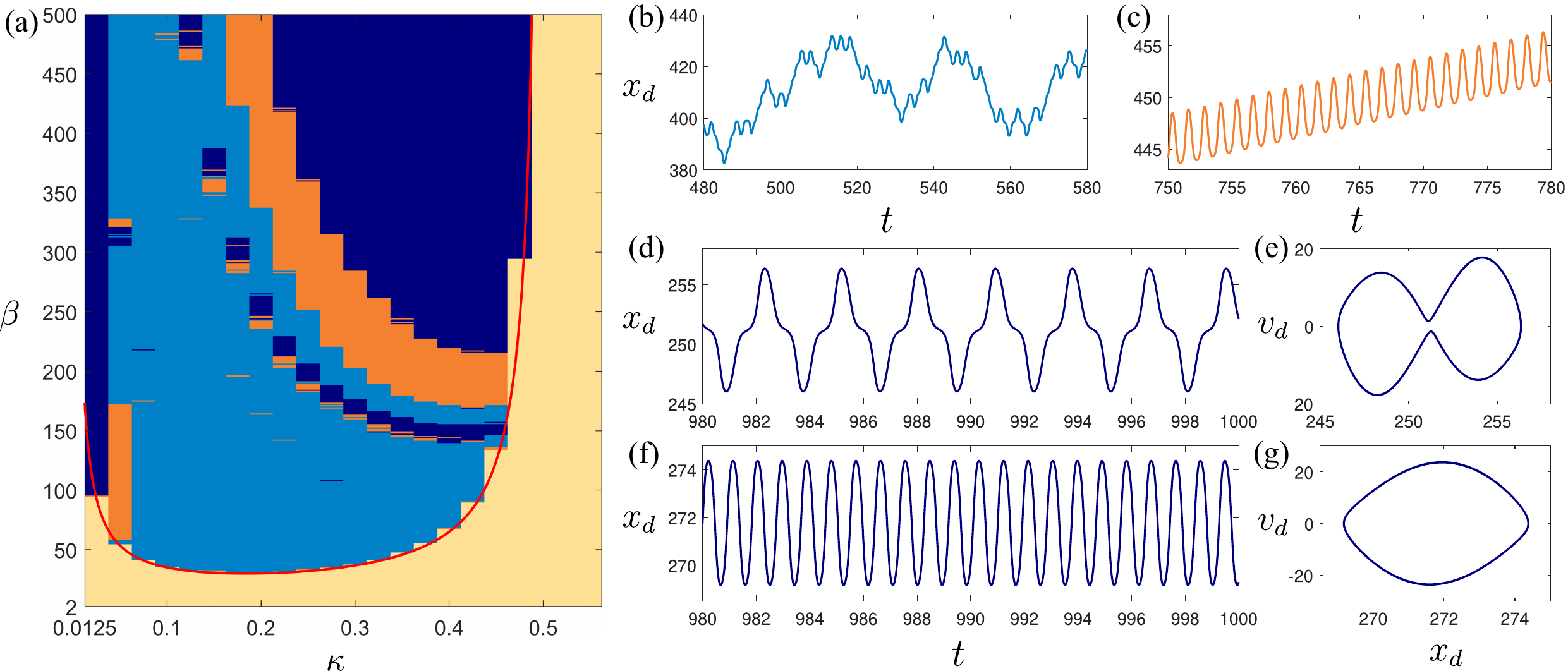}
\caption{Walking behaviors for a sinusoidal wave form.  (a) Different dynamical behaviors observed in the $(\kappa,\beta)$ parameter space at $t=1000$ from simulations initiated at $t=0$ with the droplet in the steady walking state for $t\leq 0$. We explore the parameter space region $0.025\leq \kappa \leq 0.55$ and $0 < \beta \leq 500$ with resolution $\Delta \kappa=0.025$ and $\Delta \beta=1$. We observe steady walking (beige), oscillating walking (orange), self-trapped oscillations (navy blue) and irregular walking (blue). The solid red curve is the linear stability curve separating the steady walking and the unsteady walking regime. Typical trajectories of (b) irregular walking ($\kappa=0.30$, $\beta=71$), (c) oscillating walking ($\kappa=0.30$, $\beta=221$), and two different kinds of  self-trapped oscillations (d) ($\kappa=0.30$, $\beta=171$) and (f) ($\kappa=0.30$, $\beta=401$) are shown. Phase space trajectories of self-trapped oscillations in (d) and (f) are shown in (e) and (g) respectively.}
\label{fig: PS_space_sin}
\end{figure*}

By simulating in the $(\kappa,\beta)$ parameter space using the sinusoidal wave form (see Appendix \ref{numericsin} for details of the numerical implementation), we observe different unsteady regimes as shown in Fig.~\ref{fig: PS_space_sin}. The three distinct unsteady behaviors identified in simulations with the Bessel wave form are also realized with the sinusoidal wave form, however, the region spanned by each of those behaviors changes significantly. The oscillating walkers and the self-trapped-oscillation states occupy a significantly larger region in the parameter space compared to the small isolated regions identified using the Bessel wave form. This is likely due to the absence of spatial decay in the sinusoidal wave field compared to the Bessel wave field, which enhances interference of the waves. We find two distinct types of self-trapped oscillations in the parameter space. Inside the unsteady lobe region for large $\kappa$ and large $\beta$, or very small $\kappa$, the self-trapped oscillations form a simple closed loop in the phase space, as shown in Fig.~\ref{fig: PS_space_sin}(g). In the other narrow self-trapped oscillation region, we find a dumbbell-shaped closed loop in the phase space as shown in Fig.~\ref{fig: PS_space_sin}(e). {We note that irregular and oscillating walkers were also observed by \citet{Durey2020lorenz}, for a sinusoidal wave form using a different dimensionless form for the droplet's equation of motion. Moreover, as in \citet{Durey2020lorenz}, we also observe hysteresis near the boundary separating steady and unsteady walking.} 

Since the equation of motion with a sinusoidal wave form is simpler than with the Bessel wave form, we explore the chaotic and statistical aspects of irregular walking mainly using the sinusoidal wave form in Secs. \ref{Dynamics irr} and \ref{Statistics irr} respectively.

\section{Dynamics in the irregular walking regime}\label{Dynamics irr}

\begin{figure*}
\centering
\includegraphics[width=2\columnwidth]{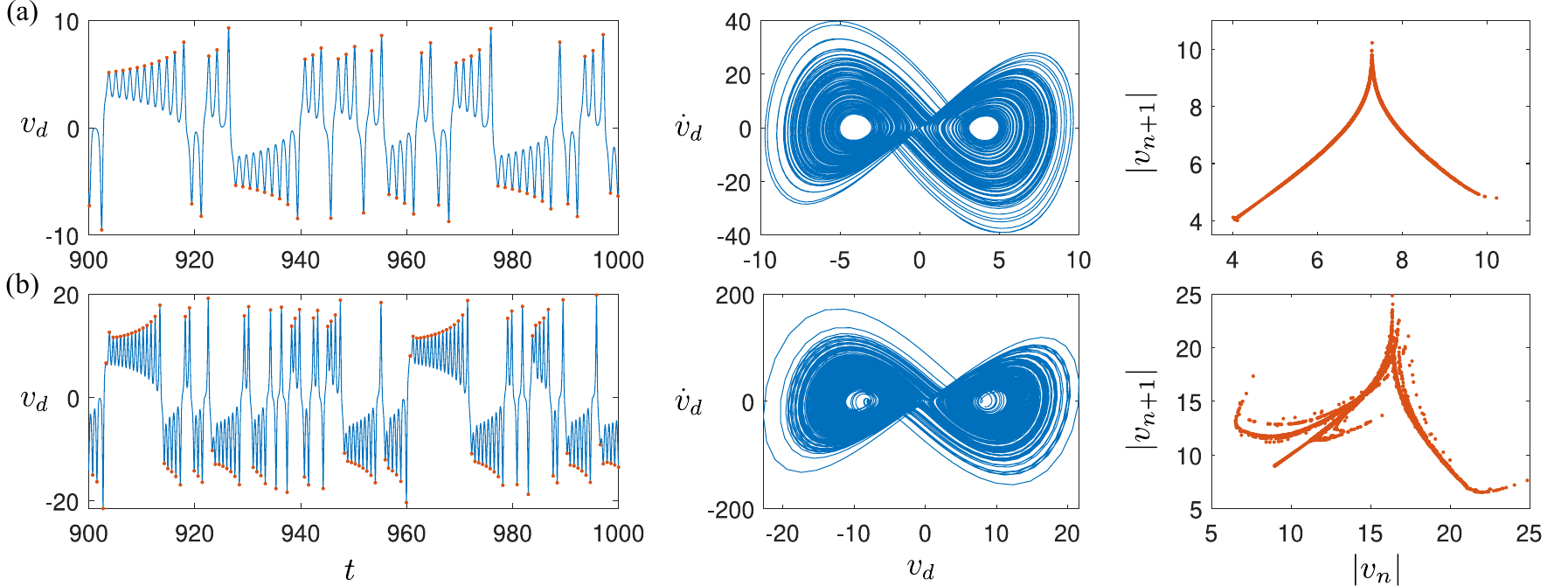}
\caption{Comparison of the chaotic behavior in the droplet's dynamics using (a) a sinusoidal wave field ($\kappa=0.2$, $\beta=35$) and (b) a Bessel wave field ($\kappa=0.1$, $\beta=90$). For the droplet's dynamics, the time series of velocity $v_d$ is shown in the left panel, the projection of the chaotic attractor in the $(v_d,\dot{v}_d)$ phase space in the middle panel and the 1D return map for the maximum speed in each oscillation in the right panel.}
\label{fig: chaos}
\end{figure*}

In the irregular walking regime of the $(\kappa,\beta)$ parameter space, we observe that the position-time trajectory of the droplet resembles a random-walk-like motion for both the Bessel and the sinusoidal wave form. To explore this in more detail, we investigate the velocity time series of the droplet undergoing irregular walking. The velocity time series, the phase space dynamics in $(v_d,\dot{v}_d)$ space and the 1D return map of the maxima of the absolute velocity for typical parameter values for a sinusoidal wave form and a Bessel wave form are shown in Figs.~\ref{fig: chaos}(a) and (b) respectively. The velocity time series has two distinct features: (i) oscillations that correspond to speed oscillations in the walking direction and (ii) flip-flop behavior that corresponds to the reversal of the walking direction. The projection of the dynamics into the $(v_d,\dot{v}_d)$ phase plane reveals the underlying chaotic attractor. A plot of the maximum speed $|v_{n+1}|$ on oscillation $n+1$ versus the maximum speed $|v_n|$ on the previous oscillation $n$ generates a cusp map~\citep{beck_schogl_1993,Balakrishnan2020}. {A discussion comparing the chaotic dynamics of the sinusoidal and Bessel pilot-wave systems can be found in the supplemental material of \citet{Durey2020lorenz}.} 

\begin{figure*}
\centering
\includegraphics[width=1.8\columnwidth]{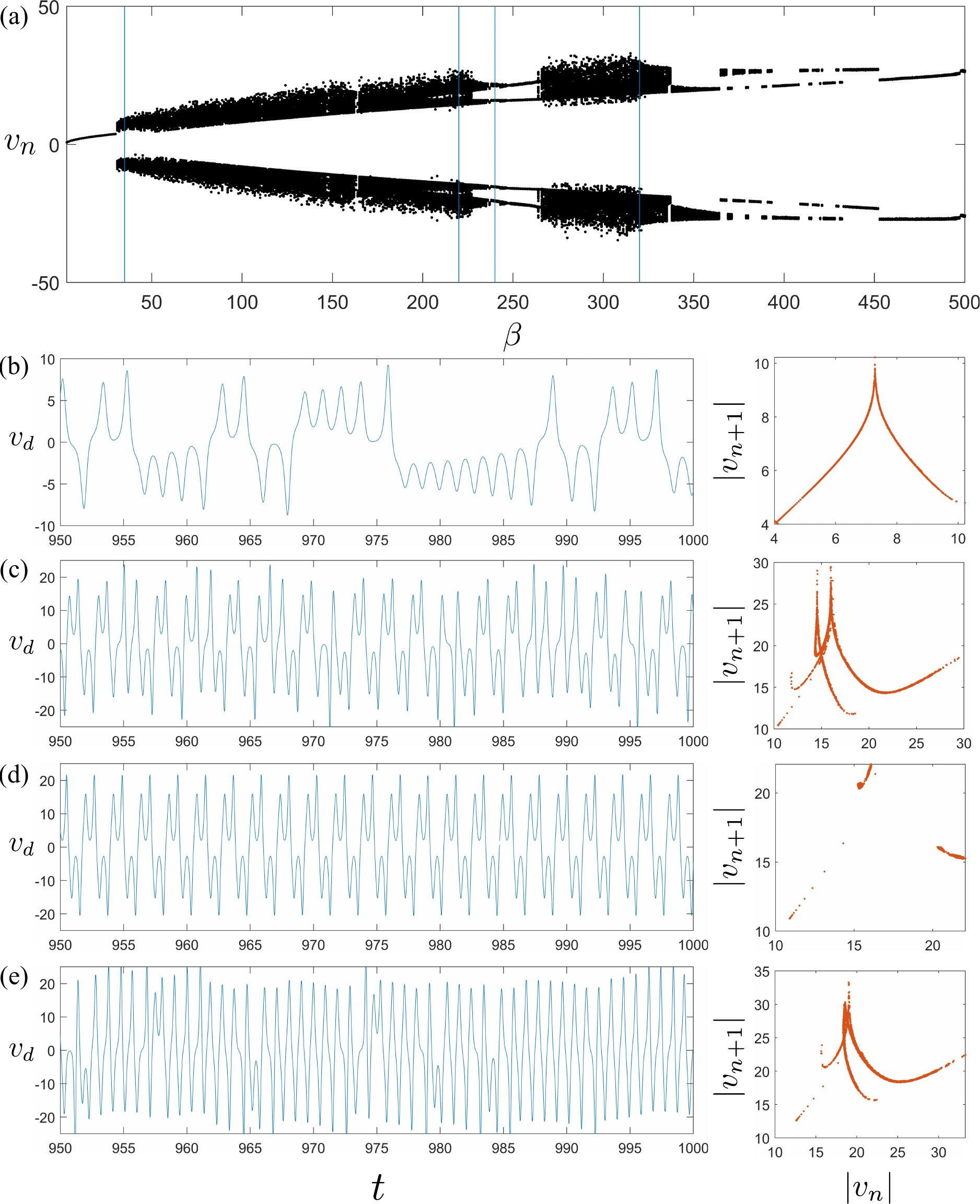}
\caption{(a) Velocity bifurcation diagram for the sinusoidal wave form showing the maxima and minima $v_n$ of the velocity time series as a function of the parameter $\beta$ and a fixed $\kappa=0.2$. The time series of velocity and the corresponding map for the maximum consecutive absolute values are shown for (b) $\beta=35$, (c) $\beta=220$, (d) $\beta=240$, and (e) $\beta=320$.}
\label{fig: chaos3}
\end{figure*}

To understand how the droplet's dynamics change in the parameter space, we have explored the velocity time series using a sinusoidal wave form as a function of the parameter $\beta$, by fixing $\kappa$. Figure~\ref{fig: chaos3}(a) shows the velocity bifurcation diagram where the maxima and minima $v_n$ in the velocity time series are plotted against the parameter $\beta$, revealing regions of periodic and chaotic dynamics. Figure~\ref{fig: chaos3}(b)-(e) shows the velocity time series and the 1D map of consecutive speed maxima at different $\beta$ values. At low $\beta$, near the onset of the unsteady regime, we find that this map has a single cusp-like structure. At large $\beta$, we see multiple cusps emerging in the multi-valued map. Interspersed between the chaotic regimes are periodic regimes where the 1D map collapses to a compact region. For $\beta\gtrsim 330$, the droplet transitions from the chaotic regime to the oscillating-walker regime and remains in the oscillating-walker state till $\beta=500$.

The structures of the chaotic attractor and the cusp map for the sinusoidal wave form in Fig.~\ref{fig: chaos}(b) have striking resemblance to the attractor and the 1D return map of the Lorenz system~\citep{Sparrowbook}. Inspired by these similarities, we explore the connection between the two systems in the next section.

\subsection{Connection to the Lorenz system}

One of the classic systems that exhibits chaotic behavior is the celebrated Lorenz system~\citep{Lorenz1963} defined as follows:
\begin{align}\label{eq: lorenz ode}
&\frac{d X}{dt}=\sigma(Y-X),\nonumber\\
&\frac{d Y}{dt}=-XZ+rX-Y,\\
&\frac{d Z}{dt}=XY-bZ.\nonumber
\end{align}
 
This system has three fixed points: (i) $X=Y=Z=0$ (unstable), (ii) $X=Y=\sqrt{b(r-1)}$ and $Z=r-1$ (stable), and (iii) $X=Y=-\sqrt{b(r-1)}$ and $Z=r-1$ (stable) for $1<r<r_c$ with $r_c=\sigma(\sigma+b+3)/(\sigma-b-1)$. The parameters $\sigma$, $r$ and $b$ are positive. When $r>r_c$, all fixed points are unstable and the system exhibits either periodic or chaotic behavior on a strange attractor~\citep{Sparrowbook}.

\citet{Takeyama1978_1} showed that the system of Lorenz equations in \eqref{eq: lorenz ode} can be recast into an integro-differential equation for the variable $X$. By eliminating the variable $Y$ in the system of ordinary differential equations (ODEs) in Eq.~\eqref{eq: lorenz ode}, we get
\begin{gather}
    \ddot{X}+(1+\sigma)X+\sigma(1-r+Z)X=0, \label{Xeq} \\
\dot{Z}+bZ=X(X+\dot{X}/\sigma). \label{Zeq}
\end{gather}

We can further eliminate $Z$ by solving Eq.~\eqref{Zeq} for $Z(t)$ and then substituting the solution into Eq.~\eqref{Xeq}. This results in the integro-differential equation
\begin{align}\label{Xfeq}
&\ddot{X}+(1+\sigma)\dot{X}+\sigma X \Big[1-r+\frac{1}{2\sigma}X^2 \nonumber \\
&+ \left(1-\frac{b}{2\sigma}\right) \int_0^{\infty} X^2(t-z) \,\text{e}^{-bz}\,\text{d}z \Big]=0. 
\end{align}

In Eq.~\eqref{Xfeq}, we have dropped the terms due to the initial values that decay exponentially in time and assumed that the motion started at an infinite time in the past~\citep{Takeyama1978_1,Festa_2002}. If we assume that $b \gg 1$ and approximate the exponential term $\text{e}^{-bz}$ in the integral of Eq.~\eqref{Xfeq} by a delta function $\delta(bz)$, then the equation reduces to
 \begin{equation}\label{eq: lorenz low memory}
\ddot{X}+(1+\sigma)\dot{X} +\frac{d U}{d X}=0, 
\end{equation}
with
 \begin{equation*}
U(X)=\sigma\left(\frac{1-r}{2}X^2+\frac{1}{4b}X^4\right).
\end{equation*}

Equation~\eqref{eq: lorenz low memory} can be interpreted as one-dimensional motion of a particle with unit mass in a quartic potential well $U(X)$ with friction coefficient $1+\sigma$~\citep{Takeyama1978_1,Takeyama1980_2}. For $r>1$, the quartic potential well takes the form of a double-well potential with stable fixed points at $X=\pm\sqrt{b(r-1)}$ and an unstable fixed point at $X=0$. The general expression in Eq.~\eqref{Xfeq} can be rewritten by splitting the $\text{e}^{-bz}$ terms into a delta function and the deviation from it, giving
\begin{align}\label{Xfeq2}
&\ddot{X}+(1+\sigma)\dot{X}+\frac{d U}{d X}\\\nonumber
&+\left(\sigma-\frac{b}{2}\right) X \int_0^{\infty} \left(X^2(t-z) - X^2(t)\right)\text{e}^{-bz}\,\text{d}z=0. 
\end{align}

The above equation can be interpreted as a particle of unit mass and a friction coefficient $1+\sigma$ in a potential well $U(X)$ with an additional force that depends on the history of the motion. Without the memory term, the particle would stop in one of the minima of the double-well potential $U(x)$, due to the damping force $-(1+\sigma)\dot{X}$. The memory forcing sustains the particle's motion. The particle oscillates in one of the minima with growing amplitude until it has sufficient energy to cross the barrier at $X=0$~\citep{Festa_2002,Festa2002pre}. 

{\citet{Durey2020lorenz} found similarities between the droplet's dynamical system and the Lorenz system by expressing the integro-differential trajectory equation in their pilot-wave model as a system of ODEs and comparing with the Lorenz system equations. By contrast, in this section, we show an \text{exact correspondence} between the integro-differential equation that governs the droplet's velocity and the integro-differential equation \eqref{Xfeq} of the Lorenz system.} 

Since the velocity of the droplet in the irregular walking regime has a chaotic attractor and a cusp map similar to the Lorenz system, let us rewrite the integro-differential equation describing the motion of the droplet in Eq.~\eqref{eq_1} in terms of the velocity variable as
\begin{equation}\label{eq: vel1}
    \kappa \dot{v}_d + v_d =\beta \int_{-\infty}^{t} f(x_d(t)-x_d(s)) \,\text{e}^{-(t-s)}\,\text{d}s.
\end{equation}
By differentiating this equation with respect to time, we obtain the integro-differential equation 
\begin{equation}\label{dropveleq}
    \kappa \ddot{v}_d +(1+\kappa)\dot{v}_d +v_d=\beta \left(f(0) + v_d M(t)\right),
\end{equation}
for the velocity of the droplet, where the memory forcing term is given by
\begin{equation*}
    M(t)=\int_{-\infty}^{t} f'\left(x_d(t)-x_d(s)\right) \text{e}^{-(t-s)}\,\text{d}s.
\end{equation*}
From the stationary solution, we know that $f(0)=0$. Hence, differentiating the above memory term with respect to time, we get 
\begin{equation*}
    \dot{M}(t)=f'(0)+v_d\int_{-\infty}^{t} f''\left(x_d(t)-x_d(s)\right) \text{e}^{-(t-s)}\,\text{d}s - M(t).
\end{equation*}
For the sinusoidal wave field, $f'(0)=1/2$ and $f''\left(x_d(t)-x_d(s)\right)=-f\left(x_d(t)-x_d(s)\right)$. Using this in combination with Eq.~\eqref{eq: vel1} we arrive at,
\begin{align*}
\int_{-\infty}^{t} & f''\left(x_d(t)-x_d(s)\right) \text{e}^{-(t-s)}\,\text{d}s\\ \nonumber
&= -\int_{-\infty}^{t} f\left(x_d(t)-x_d(s)\right) \text{e}^{-(t-s)}\,\text{d}s\\ \nonumber 
&= -\frac{1}{\beta}(\kappa \dot{v}_d+v_d).
\end{align*}
Hence, we get
\begin{equation*}
    \dot{M}(t)+M(t)=\frac{1}{2}-\frac{v_d}{\beta} (\kappa \dot{v}_d+v_d).
\end{equation*}
Solving this ODE for $M(t)$ gives
\begin{align*}
    M(t)&=\int_{-\infty}^{t} \left(\frac{1}{2}-\frac{v_d}{\beta} (\kappa \dot{v}_d+v_d)\right) \text{e}^{-(t-s)} \text{d}s\\\nonumber
    &=\frac{1}{2}-\frac{\kappa}{2\beta}v_d^2+\frac{\kappa-2}{2\beta}\int_{-\infty}^{t} v_d^2(s) \,\text{e}^{-(t-s)} \text{d}s.
\end{align*}
Finally, substituting this into Eq.~\eqref{dropveleq}, we get
\begin{align}\label{dropveleqF}
&\ddot{v}_d +\left(1+\frac{1}{\kappa}\right)\dot{v}_d\\\nonumber &+\frac{v_d}{\kappa}\left[1-\frac{\beta}{2}+\frac{\kappa}{2} v_d^2+\frac{2-\kappa}{2}\int_{0}^{\infty} v_d^2(t-z) \text{e}^{-z} \text{d}z\right]=0.    
\end{align}

By comparing Eqs.~\eqref{Xfeq} and \eqref{dropveleqF}, we find an exact correspondence, with the parameters in the droplet system related to the Lorenz system via
\begin{equation*}
    b=1,\,r=\frac{\beta}{2}\,\text{ and }\,\sigma=\frac{1}{\kappa}.
\end{equation*}

Hence, using a sinusoidal wave form in the walker system, the droplet's velocity $v_d(t)$ is equivalent to the variable $X(t)$ in the Lorenz system. Thus, similarly to the interpretation of the variable $X$ in the Lorenz system, one may interpret the velocity variable $v_d$ in the above droplet's integro-differential equation as the position of a fictitious particle of unit mass and a friction coefficient $1+1/\kappa$ in a double-well potential $U(v_d)$, with an additional force that depends on the history of the motion. Conversely, one may also interpret the variable $X$ in the Lorenz system as the velocity of a droplet of mass $1/\sigma$ which is subject to a drag force $-X$ and propelled by the memory force from the underlying sinusoidal wave that the droplet generates continuously. We note that chaotic attractors that are qualitatively similar to the Lorenz system have also been reported in simulations of a walker with a Bessel wave form in a central harmonic potential~\citep{doi:10.1063/1.5058279}.

We have shown the precise sequence of transformations that map the droplet dynamics to those of the Lorenz system in the case of a sinusoidal wave form. As such, our droplet dynamics immediately inherit the rich array of features that the Lorenz system possesses, including but not limited to chaotic dynamics, invariant manifold theorems, and bifurcations. The invariant manifolds of the Lorenz system are notoriously difficult to compute but can be used to understand the chaotic dynamics. For instance, it has been shown that the 2D stable manifold of the fixed point at the origin is a phase space separatrix, with all trajectories (including those on the butterfly wings) sandwiched between the sheets of this manifold \cite{Osinga2002,Osinga2018}. Supported by the results of our simulations, we expect that these invariant manifolds and the roles they play in organizing the phase space, will persist when the oscillations in the wave form are modulated, as in the Bessel wave form.

\begin{figure*}[!t]
\centering
\includegraphics[width=2\columnwidth]{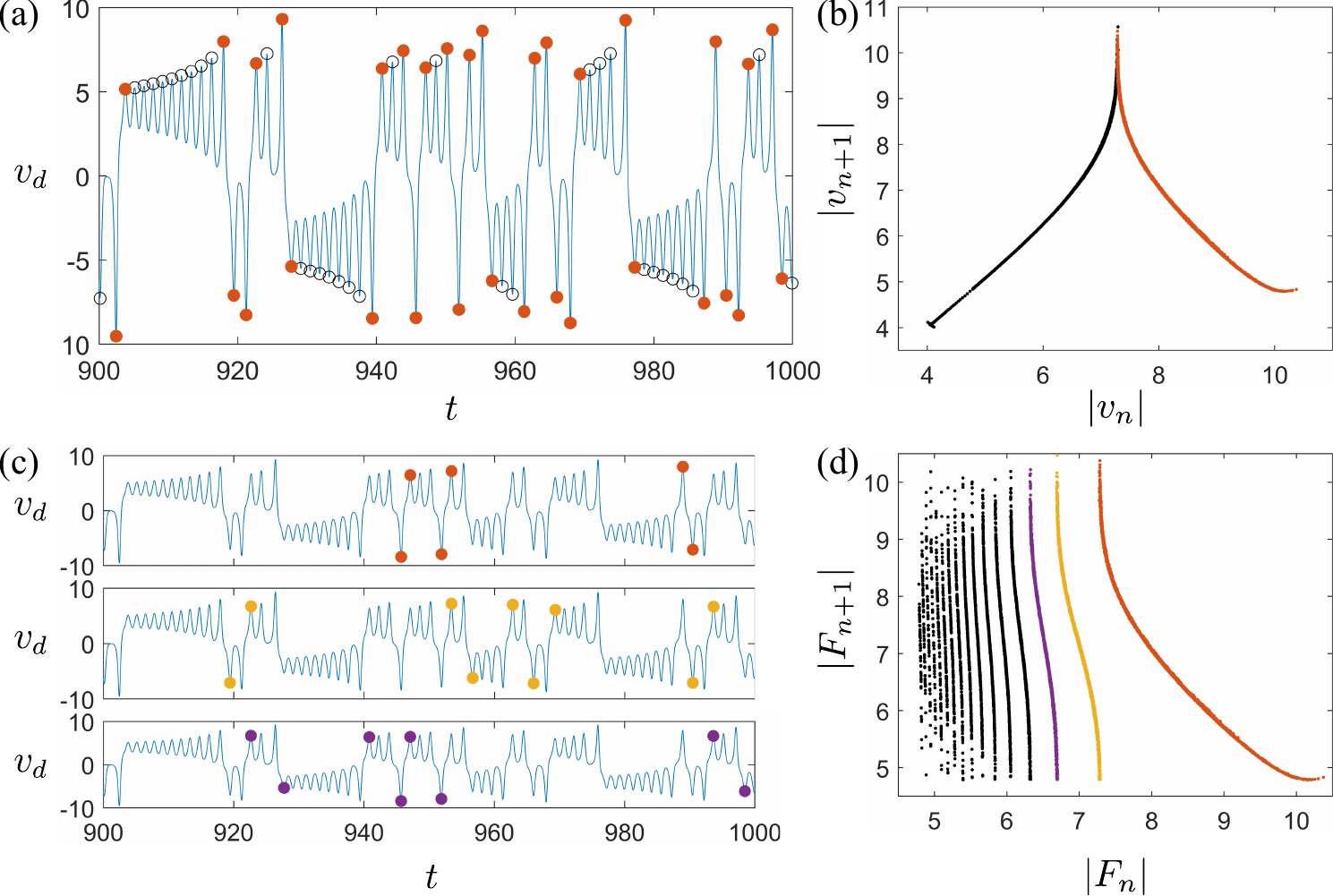}
\caption{Switching dynamics in the irregular walking regime for the sinusoidal wave form at $\kappa=0.2$ and $\beta=35$. (a) The velocity time series along with circles at the extrema of the oscillations are shown. The red filled circles indicate the extreme values of velocity before and after the flip. (b) Return map of absolute value of consecutive extrema in the time series, i.e. the absolute value of the consecutive circles in the left panel. The red and black branch corresponds to the red and black circles in the time series. (c) The same time series as in (a) but the markers now highlight the extreme values after a flip with a fixed number of $N$ oscillations between them. The $N=0$ (red circles), $N=1$ (yellow circles), and $N=2$ (purple circles) are shown. (d) The map showing consecutive absolute values of the extrema after a flip with the $N=0$, $N=1$, and $N=2$ branch highlighted.}
\label{fig: chaos2}
\end{figure*}


\subsection{Switching dynamics of irregular walking}

{\citet{Durey2020lorenz} made a connection between the particle's irregular switching dynamics in the sinusoidal model and the 1D cusp like map that arises by plotting consecutive maxima of the local wave amplitude. Here we analyze this connection by producing maps of consecutive maxima in speed oscillations.} As shown in Figs.~\ref{fig: chaos2}(a) and (b), we find that the ascending branch on the cusp map corresponds to maxima in speed oscillations when the droplet is moving in a given direction (black empty circles) while the descending branch corresponds to the maxima in speed oscillations when a flip occurs in the velocity time series or equivalently a reversal in the walking direction (red filled circles). 

To analyze the particle's switching dynamics further, we look at the maximum absolute velocity after a flip occurs with a fixed number $N$ of oscillations between the consecutive flips and plot these consecutive values against each other (see Figs.~\ref{fig: chaos2}(c) and (d)). This map of consecutive values results in a band-like structure similar to the Continued Fraction map, also known as the Gauss map~\citep{Gaussmap,Balakrishnan2020}. We find that in this map, each band corresponds to a fixed number $N$ of oscillations between consecutive direction reversals. The branches corresponding to $N=0$ (red), $N=1$ (yellow) and $N=2$ (purple) are shown. 

\section{Statistical aspects of irregular walking}\label{Statistics irr}

We now turn to explore the statistical properties of the random-walk-like dynamics observed in the irregular walking regime with a sinusoidal wave form. 

\begin{figure*}
\centering
\includegraphics[width=2\columnwidth]{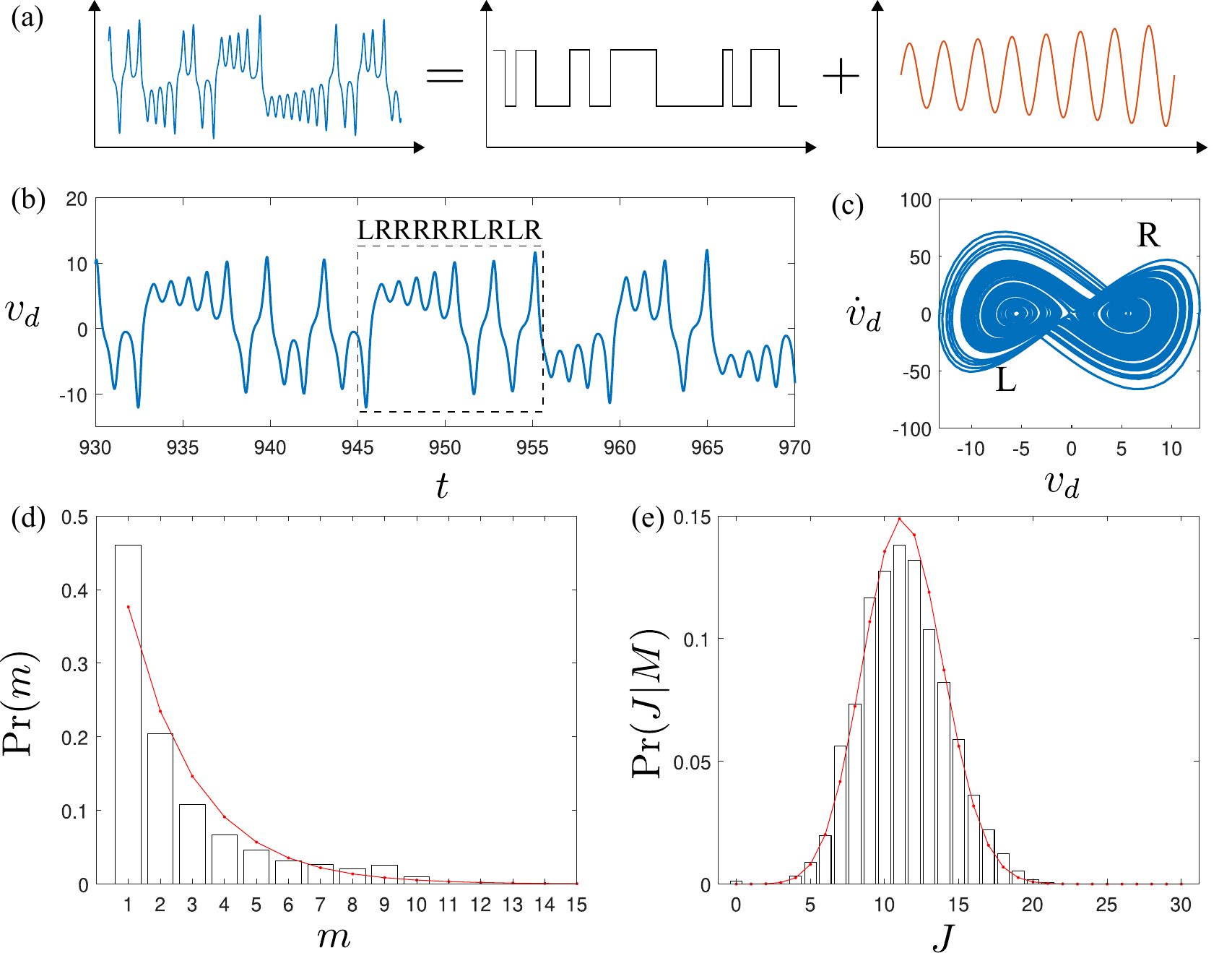}
\caption{Statistics of the flip-flop process for $\kappa=0.2$ and $\beta=65$ using the sinusoidal wave field. (a) Schematic showing that a typical velocity time-series of the droplet in the irregular walking regime can be thought of as a sum of a flip-flop process and an exponentially increasing sinusoid which dictates the flip after the amplitude reaches some threshold value~\citep{Aizawa1982}. (b) Time series of velocity for a typical droplet's trajectory in the irregular walking regime and (c) the corresponding projection of the chaotic attractor. The attractor has two basins that are labeled left `$L$' and right `$R$'. (d) Probability distribution for having $m$ oscillations between flips. (e) Probability distribution for the number of jumps $J$ in a given sequence of $M=30$ steps. In both (d) and (e), the histogram is from the numerical simulations while the red curves are best fits obtained using Eqs.~\eqref{Eq: Prob dist 1} and \eqref{Eq: Prob dist 2} respectively.}
\label{fig: statchaos1}
\end{figure*}

\subsection{Statistical properties of irregular switching of walking direction}

\citet{Aizawa1982} analyzed the statistical aspects of the Lorenz system by decomposing the time series for the  system into a flip-flop process and sinusoidal oscillations with increasing amplitude (see Fig.~\ref{fig: statchaos1}(a)). We take a similar approach for the droplet's velocity time series and focus on the statistical aspects of the flip-flop process that dictates reversals in the walking direction.

The flip-flop process can also be thought of as the switches between the two attracting basins of the chaotic attractor shown in the middle panel of Figs.~\ref{fig: chaos}(a) and (b). Denoting the left and right attracting basins by $L$ and $R$ respectively, the dynamics of the flip-flop process will generate a sequence of states $LLRRLR...$ for each trajectory (see Figs.~\ref{fig: statchaos1}(b) and (c)). The probability of being found in each state, $L$ or $R$, is given by $\text{Pr}(L)=\text{Pr}(R)=1/2$, due to the symmetry of the system. We investigate the statistics of the flip-flop process and compare it to a Markovian process. For a Markovian flip-flop process, the transition probabilities are constant and hence calling $p$ the probability of flipping or reversing the walking direction, we have $\text{Pr}(L|R)=\text{Pr}(R|L)=p$, while the probability of maintaining the walking direction is given by $\text{Pr}(L|L)=\text{Pr}(R|R)=1-p$~\citep{Aicardi1987}.

For a sequence $LLRLRRLRRL...$, we can generate a chain $NJJJNJJNJ...$, where $J$ denotes an occurrence of a jump or a walking direction reversal and $N$ denotes that no jump has occurred. If the process is Markovian, then the probability that the phase-space trajectory will execute $m$ turns after entering a basin before it jumps out of the basin is given by the geometric distribution~\citep{Aizawa1982,Aicardi1987}
\begin{equation}\label{Eq: Prob dist 1}
    \text{Pr}(m)=p(1-p)^{m-1}.
\end{equation}
Similarly, for a Markovian process, the probability that $J$ jumps have occurred in a sequence of $M$ turns is given by~\citep{Aizawa1982,Aicardi1987}
\begin{equation} \label{Eq: Prob dist 2}
\text{Pr}(J|M)= {M \choose J} p^J (1-p)^{M-J}.   
\end{equation}

We can estimate these probabilities from sufficiently long  chains of the flip-flop process for the walker from simulations by using
\begin{equation} \label{Eq: Prob dist}
    \text{Pr}(m)=\frac{\sum_{k=1}^{N} \delta_{m,i_k}}{N}\:\:\text{and}\,\,\text{Pr}(J|M)=\frac{\sum_{k=1}^{T} \delta_{J,J_k}}{T}.
\end{equation}
Here $\delta$ is the Kronecker delta, $i_k$ represents the number of turns executed in the basin between the $(k-1)$th and $k$th jump, $N$ is the total number of jumps in the sample and $J_k$ is the number of jumps occurring in each of the $T$ sequences of $M$ turns of the $k$th subdivided sample. 

Figures~\ref{fig: statchaos1}(d) and (e) show the probability distributions calculated from an ensemble average of $100$ trajectories in the simulations (histograms) and the corresponding best fits of Eqs.~\eqref{Eq: Prob dist 1} and \eqref{Eq: Prob dist 2} for a typical $\kappa$ and $\beta$ value in the irregular walking regime. For the parameters chosen in Fig.~\ref{fig: statchaos1}, we obtain $p=0.376$. We find a good fit for these parameters, suggesting that the Markovian approximation is reasonable in certain regions of the parameter space. However, we note that this Markovian approximation does not give a good fit in the entire irregular walking regime of the parameter space. This is also true for the Lorenz system where the Markovian approximation gives a good fit in some regions of parameter space while in other regions there are sharp deviations from the Markovian process~\cite{Aicardi1987}. {We note that \citet{Durey2020} also described the irregular walking as a flip-flop process to rationalize the Gaussian-like probability density function of the droplet's position at long times.}

\subsection{Connection with the Langevin equation}

We also compare the equation of motion of the droplet in Eq.~\eqref{eq_1}, with a Langevin-type equation that describes the motion of a particle under stochastic forcing~\citep{Sancho1984},
\begin{equation}\label{Eq: langevin}
    \dot{q}+\gamma q = \xi(t).
\end{equation}
Here $q$ is the dynamical variable, $\gamma$ is the friction coefficient and $\xi(t)$ is the stochastic forcing. Comparing this equation with the droplet's equation of motion for a sinusoidal wave field,
\begin{equation}\label{droplet equation}
\dot{v}_d+\frac{1}{\kappa}v_d
=\frac{\beta}{\kappa}\int_{0}^{\infty}\frac{1}{2}\sin(x_d(t)-x_d(t-z))\text\,\text{e}^{-z}\,\text{d}z,
\end{equation}
we see that the dynamical variable $q$ is equivalent to the velocity $v_d$ of the droplet, the friction coefficient $\gamma$ is equivalent to $1/\kappa$ and the stochastic forcing in Eq.~\eqref{Eq: langevin} takes the place of the memory forcing in Eq.~\eqref{droplet equation}. The wave force on the walker in the irregular walking regime has a time series similar to the velocity time series where the oscillating force switches erratically between positive and negative values. We can crudely approximate the memory-force time series for the droplet as a flip-flop process and ignore the oscillations. Then, the force time series of the droplet resembles a dichotomous process where the values of the force flip randomly between only two possible values. If we choose the stochastic noise $\xi(t)$ in Eq.~\eqref{Eq: langevin} to be a dichotomous process, then we can compare the Langevin dynamics with the numerical simulations of the droplet's dynamics. We assume that $\xi(t)$ is a dichotomous process that will have only two possible values $\pm \Delta$ with equal probability and jumps between them at a rate $\lambda/2$~\citep{Sancho1984}. This form of the forcing has zero mean and autocorrelation
\begin{equation*}
    \langle\xi(t)\xi(t')\rangle=\Delta^2 \text{e}^{-\lambda |t-t'|}.
\end{equation*}

\begin{figure}
\centering
\includegraphics[width=0.95\columnwidth]{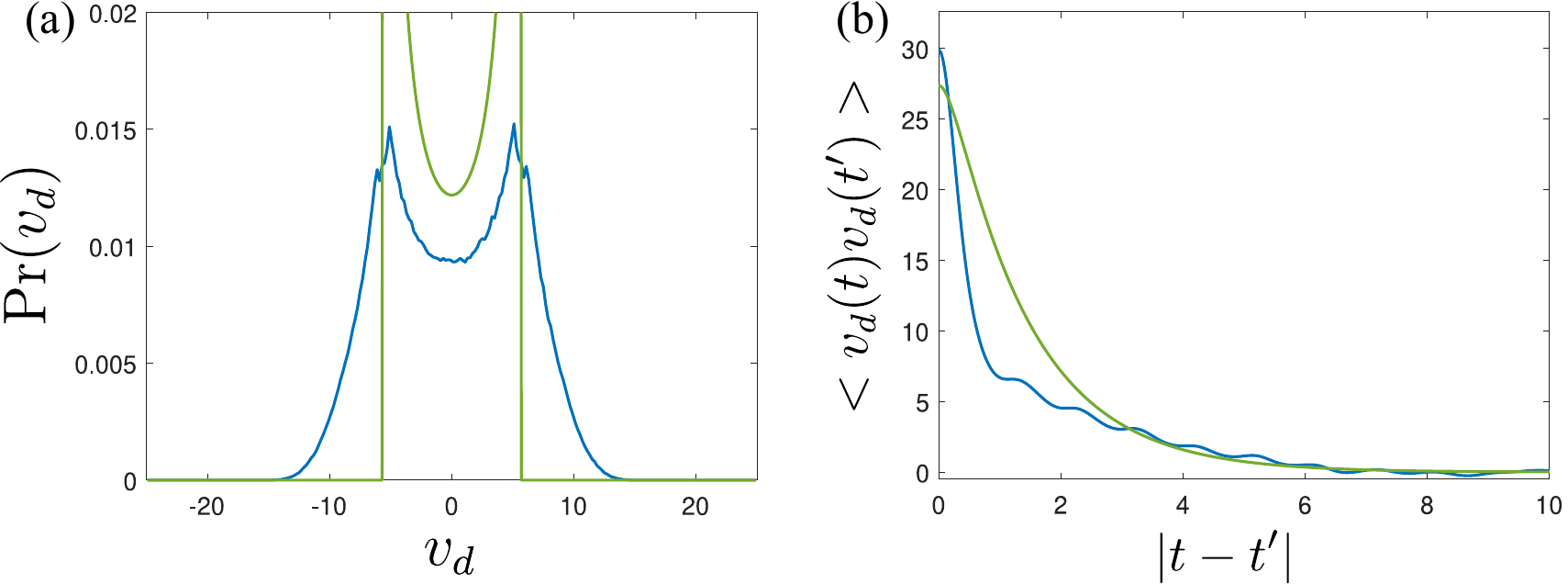}
\caption{Stationary probability distribution and autocorrelation for the velocity flip-flop process at $\kappa=0.2$ and $\beta=65$ using the sinusoidal wave field. (a) Stationary probability distribution for velocity. (b) Velocity autocorrelation function. In both panels, the blue curve is from numerical simulations of a walker while the green curve is the fit obtained by using the Langevin model with dichotomous noise. The steady walking speed $u=\sqrt{\beta/2 - 1}=5.6125$ is very close to the boundary of the Langevin model (vertical green line) in (a).}
\label{fig: statchaos2}
\end{figure}

\begin{figure*}[!t]
\centering
\includegraphics[width=1.8\columnwidth]{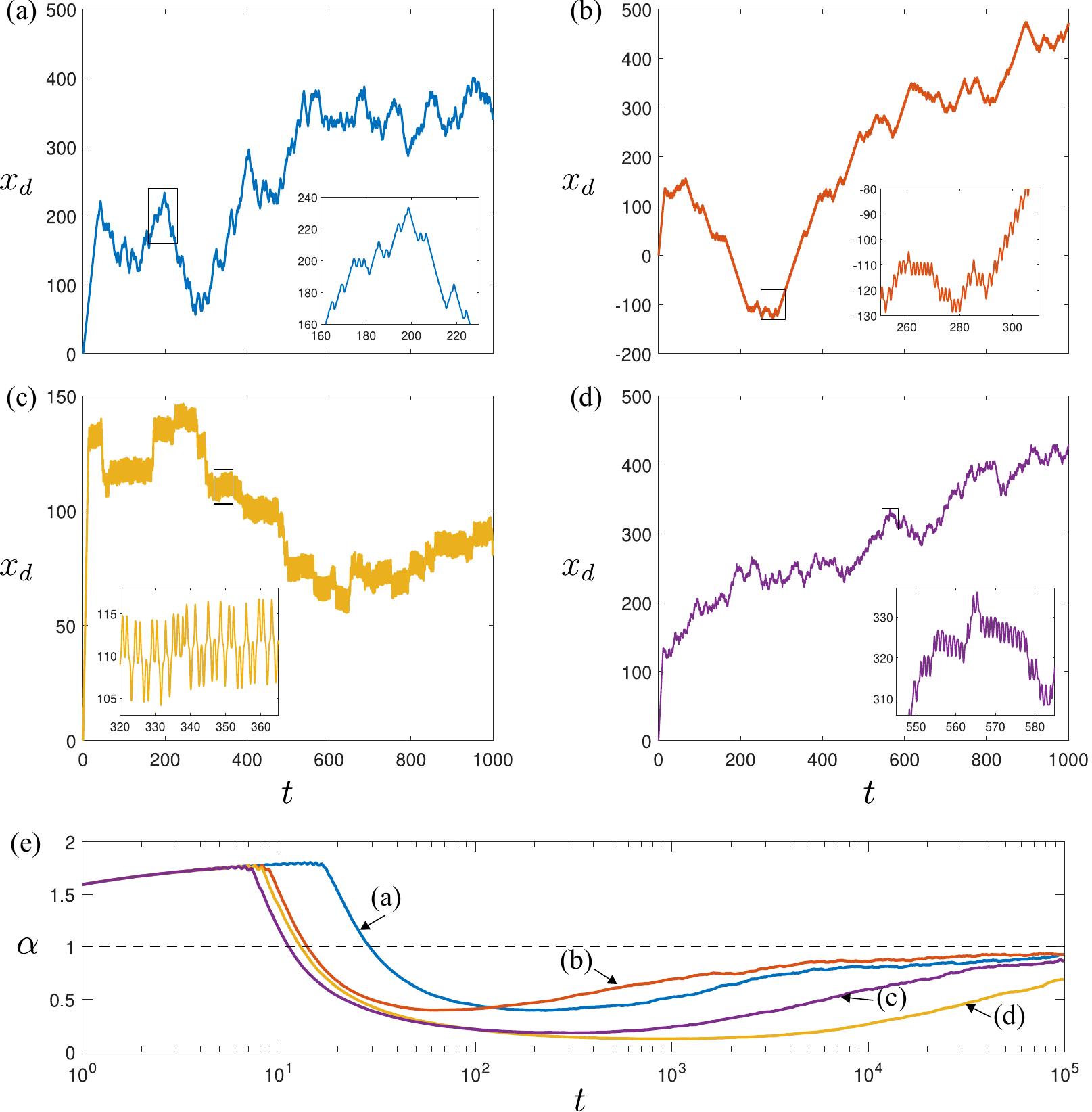}
\caption{Diffusion-like motion of the droplet in the irregular walking regime for the sinusoidal wave form. Typical trajectories at (a) $\beta=65$, (b) $\beta=165$, (c) $\beta=200$, and (d) $\beta=300$ are shown for a fixed $\kappa=0.2$. (e) Time dependent diffusion exponent $\alpha(t)$ as a function of time for the trajectories shown in (a)-(d). For the statistical analysis, an ensemble of $5000$ trajectories were simulated with each trajectory run for $t=10^5$ from which we sampled $1000$ logarithmically spaced points. The $5000$ trajectories were generated by adding a small random perturbation in the range $[0,0.1]$ in the droplet's steady walking speed at the start of the simulation.}
\label{fig: Diff_dynamics}
\end{figure*}

For the droplet's dynamics this value of $\Delta$ can be approximated by $\Delta\approx u/\kappa$ where $u=\sqrt{\beta/2 - 1}$ is the steady walking speed for the sinusoidal wave form. For the Langevin equation described in Eq.~\eqref{Eq: langevin} with a dichotomous noise term, the exact solution for the stationary probability distribution of the variable $q$  is~\citep{Sancho1984}
\begin{equation}
P_{st}(q)=N(\Delta^2 - \gamma^2 q^2)^{\lambda/2\gamma-1},    
\end{equation}
where
\begin{equation}
    N=\frac{\gamma\Gamma(1/2+\lambda/2\gamma)}{\Delta^{\lambda/\gamma-1}\Gamma(1/2)\Gamma(\lambda/2\gamma)}.
\end{equation}
The corresponding autocorrelation function 
is
\begin{equation}
\langle q(t)q(t') \rangle =\frac{-\lambda \Delta^2}{\gamma(\gamma^2-\lambda^2)}\text{e}^{-\gamma(t-t')} + \frac{\Delta^2}{\gamma^2-\lambda^2}\text{e}^{-\lambda(t-t')}.    
\end{equation}

A comparison of the Langevin model results with the numerical results for the stationary probability distribution of the droplet's velocity and velocity autocorrelation function is shown in Figs.~\ref{fig: statchaos2}(a) and (b) respectively. We find that the Langevin model captures the qualitative features of both of these plots. {We note that \citet{durey2018} also used a Langevin equation formalism to rationalize the wave-like statistics emerging at long times when the droplet is confined in a harmonic potential. Moreover, \citet{Hubert2019} also used velocity autocorrelation functions to describe the statistical properties of the long term dynamics in the irregular walking regime of a free walker in two dimensions.}

\subsection{Analysis of the diffusion-like dynamics}


In 1828, Robert Brown observed the erratic motion of small particles suspended in water~\citep{Brown1828,Brown1829}. We now know this as Brownian motion. Brownian motion plays a key role in modeling many random behaviors in nature and is typically modeled by considering random impulsive forces acting on particles. However, numerous investigations have shown the existence of Brownian-like motion from deterministic dynamics (also known as deterministic diffusion) in both discrete and continuous systems (see \citep{BECK1996419,SHIMIZU1993113,CHEW2002275,Trefan1992,Festa_2002,HUERTACUELLAR20142740} and the references therein). In particular, deterministic diffusion has been shown in differential delay equations~\citep{Festa_2002,Lei2011}. 

By investigating the trajectories in the irregular walking regime of the droplet with a sinusoidal wave form, we also obtain diffusion-like behavior for the droplet. The diffusive behavior of a system can be characterized by calculating how the mean squared displacement (MSD) scales with time, i.e., $\text{MSD}=\langle(x_d(t)-x_d(0))^2\rangle \sim t^\alpha$ with $\alpha$ being the diffusion exponent. If $\alpha=2$ then the motion is ballistic while diffusive motion has $0< \alpha < 2$ with $0< \alpha < 1$ indicating sub-diffusive behavior, $\alpha=1$ indicating `normal' diffusive motion and $1< \alpha < 2$ indicating super-diffusive behavior. Some typical trajectories of the diffusion-like motion in the irregular walking regime are shown in Figs.~\ref{fig: Diff_dynamics}(a)-(d). We observe time dependent diffusive behavior. To quantify this, we define a time dependent diffusion exponent $\alpha(t)=\text{d}(\text{log}(\text{MSD}))/\text{d}(\text{log}(t))$ and plot it as a function of time as shown in Fig.~\ref{fig: Diff_dynamics}(e). We observe sub-diffusion for the time scales where the simulations have been performed and it seems to be approaching `normal' diffusion i.e. $\alpha=1$ asymptotically. {We note that \citet{Durey2020} and \citet{Durey2020lorenz} reported asymptotic diffusion of the 1D droplet dynamics in certain parameter regimes. Moreover, \citet{Hubert2019} also reported diffusive dynamics at long times in their exploration of the bimodal erratic motion of the walker in two dimensions.}

We also note that as previously described in Eq.~\eqref{Xfeq2}, the Lorenz system can be written as an integro-differential equation that describes a particle in a double-well potential. If in this equation the double-well potential is replaced by a periodic potential, then diffusive-like behavior in the variable $X$ is observed~\citep{Festa_2002} that is similar to what we see for the droplet's position $x_d$ in Figs.~\ref{fig: Diff_dynamics} (a-d). 

\section{Discussion and Conclusion}\label{DC}

We have explored the dynamics of a particle-wave entity in the $(\kappa,\beta)$ parameter space using the stroboscopic model of \citet{Oza2013} with different wave forms. We find that the steady walking state is always stable for a Gaussian wave form above the walking threshold, while for both the Bessel and the sinusoidal wave form, the steady walking becomes unstable for large $\beta$ and small $\kappa$. By choosing a sinusoidal wave form with a Gaussian envelope and varying the length scale of spatial decay, we find that the instability region diminishes as the spatial decay is enhanced, suggesting that oscillations in the wave form are an essential dynamic feature for the instability of the steady walking motion. Moreover, the presence of even small spatial decay in the wave form results in the revival of the steady walking state for very large $\beta$. In the unsteady regime for the Bessel wave form and the sinusoidal wave form, we describe a variety of unsteady motions such as oscillating walkers, self-trapped oscillations and irregular walkers. The oscillating walkers and self-trapped oscillations span an extended region in the parameter space for a sinusoidal wave form, while the presence of spatial decay in the Bessel wave form severely contracts the region spanned by both oscillating walkers and self-trapped oscillations.

Investigation of the irregular walking regime reveals that the projected chaotic attractor in the $(v_d,\dot{v}_d)$ space has a striking resemblance to the Lorenz attractor, with the corresponding 1D return maps showing a similar cusp structure. In fact for the sinusoidal wave form, we find a one-to-one correspondence between the droplet's velocity $v_d$ and the dependent variable $X$ in the Lorenz system. \citet{Durey2020lorenz} also explored bifurcations of the droplet's dynamics using established properties of the Lorenz system and identified several pilot-wave phenomena. This suggests a deeper connection between the dynamical system underlying walkers and the Lorenz system, and warrants further investigation. We also investigated the cusp map for the sinusoidal wave form and identified the different structures in the map with the corresponding physical dynamics of the droplet.

On exploring the statistical aspects of the time series for the droplet's velocity in the irregular walking regime, we find that in certain regions of the parameter space, the statistics of the reversals in the walking direction can be well approximated by a Markovian process. Moreover, by using the Langevin equation with dichotomous noise, we are able to capture the qualitative aspects of the stationary velocity distribution as well as the velocity autocorrelation function in the droplet's dynamics.

In summary, we have made explicit connections of the walking-droplet system to the Lorenz equations, the Langevin equation and deterministic diffusion. In future, it would be interesting to investigate the unsteady droplet dynamics in the $(\kappa,\beta)$ parameter space using a stroboscopic model that allows the droplet to move in two spatial dimensions. The extra dimension may reveal novel unsteady regimes and alter the diffusive properties of the droplet dynamics.

\acknowledgements

We thank Andy Hammerlindl for useful discussions. We acknowledge financial support from an Australian Government Research Training Program (RTP) Scholarship (R.V.) and the Australian Research Council via the Future Fellowship Project No.\ FT180100020 (T.S.).

\appendix 

\section{Numerical simulations with the Bessel wave form}\label{numericsbessel}

We numerically integrate Eq.~\eqref{eq_1} with the Bessel wave form using a semi-implicit Euler method. The second-order integro-differential equation in \eqref{eq_1} can be rewritten as
\begin{equation}\label{bessel1}
\dot{x}_d=v_d,
\end{equation}
and
\begin{equation}\label{bessel2}
\dot{v}_d=\frac{1}{\kappa}\left[\beta \int_{-\infty}^{t}\text{J}_1(x_d(t)-x_d(s))\,\text{e}^{-(t-s)}\,\text{d}s - v_d\right].
\end{equation}

We assume the droplet is in the steady walking state with velocity $u$ for $t \leq 0$. Discretizing Eqs.~\eqref{bessel1} and \eqref{bessel2} using an explicit and an implicit Euler step respectively, results in
\begin{equation}\label{bessel1_dis}
x_d(t_{i+1})=x_d(t_{i})+\Delta t\, v_d(t_{i})
\end{equation}
and
\begin{align}\label{bessel2_dis}
& v_d(t_{i+1})=\frac{1}{1+\Delta t /\kappa}\Big[v_d(t_{i})+\frac{\Delta t \beta}{\kappa}\Big( I(t_{i+1}) \nonumber \\
& +\int_{0}^{t}\text{J}_1(x_d(t_{i+1})-x_d(s))\,\text{e}^{-(t_{i+1}-s)}\,\text{d}s\Big)\Big],
\end{align}
where the integral due to the initial condition $I(t_{i+1})$ is given by
\begin{equation}
    I(t_{i+1})=\int_{-\infty}^{0}\text{J}_1(x_d(t_{i+1})-u s)\,\text{e}^{-(t_{i+1}-s)}\,\text{d}s.
\end{equation}

The dimensionless time step was fixed at $\Delta t=2^{-8}$. The integral in Eq.~\eqref{bessel1_dis} was performed using the $\mathtt{MATLAB}$ trapezoid function where we considered the contribution from all the previous impacts for the first 5120 timesteps ($t=20$ using $\Delta t=2^{-8}$) and then the contributions from the last 1280 timesteps for $t>20$. At 5120 previous impacts, the exponential time damping factor reached $e^{-20} \approx 10^{-9}$ so we neglected all contribution from impacts beyond 5120 previous steps. We used an implicit step for the velocity equation because the unsteady motion of the droplet arises in the region of parameter space with very small $\kappa$ and very large $\beta$, where the integro-differential equation describing droplet motion becomes stiff. The initial-condition integral $I(t_{i+1})$ was performed using the $\mathtt{MATLAB}$  `integral' function that uses global adaptive quadrature.

\section{Numerical simulations with the sinusoidal wave form}\label{numericsin}

To simulate the droplet's dynamics for a sinusoidal wave field, we can simplify the equation of motion by changing the integro-differential equation into a finite system of ordinary differential equations (ODEs). Substituting the sinusoidal wave form in Eq.~\eqref{eq_1} and using the addition formula for sine, we obtain
\begin{align*}
\kappa\ddot{x}_d+\dot{x}_d=\frac{\beta}{2}\Big[&\sin(x_d(t))\int_{-\infty}^{t}\cos(x_d(s))\,\text{e}^{-(t-s)}\text{d}s\\\nonumber
-&\cos(x_d(t))\int_{-\infty}^{t}\sin(x_d(s))\,\text{e}^{-(t-s)}\text{d}s\Big].
\end{align*}
We define $y(t)=\int_{-\infty}^{t}\cos(x_d(s))\,\text{e}^{-(t-s)}\text{d}s$ and $z(t)=\int_{-\infty}^{t}\sin(x_d(s))\,\text{e}^{-(t-s)}\text{d}s$. These auxiliary variables satisfy
\begin{equation*}
\dot{y}+y=\cos(x_d(t))
\end{equation*}
and
\begin{equation*}
\dot{z}+z=\sin(x_d(t)).
\end{equation*}
Further letting $\dot{x}_d=v_d$, we obtain the system of ODEs~\citep{phdthesismolacek}
\begin{align} \label{sine_system_eq}
\dot{x}_d&=v_d, \\
\kappa\dot{v}_d+v_d&=\tfrac{1}{2}\beta\left[y\,\sin(x_d)-z\,\cos(x_d)\right], \\
\dot{y}+y&=\cos(x_d),\\
\dot{z}+z&=\sin(x_d).
\end{align}
We solve the system of Eqs.~\eqref{sine_system_eq} in $\mathtt{MATLAB}$  using the inbuilt ode45 solver. We initialized the simulations with the droplet in the steady walking state for $t \leq 0$, which results in the following initial conditions for the system of ODEs: $x_d(0)=0$, $v_d(0)=u$, $y(0)=1/(1+u^2)$ and $z(0)=-u/(1+u^2)$. The simulations were run for a time $t=1000$.

\bibliography{Unsteady_droplet}

\begin{thebibliography}{68}%
\makeatletter
\providecommand \@ifxundefined [1]{%
 \@ifx{#1\undefined}
}%
\providecommand \@ifnum [1]{%
 \ifnum #1\expandafter \@firstoftwo
 \else \expandafter \@secondoftwo
 \fi
}%
\providecommand \@ifx [1]{%
 \ifx #1\expandafter \@firstoftwo
 \else \expandafter \@secondoftwo
 \fi
}%
\providecommand \natexlab [1]{#1}%
\providecommand \enquote  [1]{``#1''}%
\providecommand \bibnamefont  [1]{#1}%
\providecommand \bibfnamefont [1]{#1}%
\providecommand \citenamefont [1]{#1}%
\providecommand \href@noop [0]{\@secondoftwo}%
\providecommand \href [0]{\begingroup \@sanitize@url \@href}%
\providecommand \@href[1]{\@@startlink{#1}\@@href}%
\providecommand \@@href[1]{\endgroup#1\@@endlink}%
\providecommand \@sanitize@url [0]{\catcode `\\12\catcode `\$12\catcode
  `\&12\catcode `\#12\catcode `\^12\catcode `\_12\catcode `\%12\relax}%
\providecommand \@@startlink[1]{}%
\providecommand \@@endlink[0]{}%
\providecommand \url  [0]{\begingroup\@sanitize@url \@url }%
\providecommand \@url [1]{\endgroup\@href {#1}{\urlprefix }}%
\providecommand \urlprefix  [0]{URL }%
\providecommand \Eprint [0]{\href }%
\providecommand \doibase [0]{https://doi.org/}%
\providecommand \selectlanguage [0]{\@gobble}%
\providecommand \bibinfo  [0]{\@secondoftwo}%
\providecommand \bibfield  [0]{\@secondoftwo}%
\providecommand \translation [1]{[#1]}%
\providecommand \BibitemOpen [0]{}%
\providecommand \bibitemStop [0]{}%
\providecommand \bibitemNoStop [0]{.\EOS\space}%
\providecommand \EOS [0]{\spacefactor3000\relax}%
\providecommand \BibitemShut  [1]{\csname bibitem#1\endcsname}%
\let\auto@bib@innerbib\@empty
\bibitem [{\citenamefont {Couder}\ \emph
  {et~al.}(2005{\natexlab{a}})\citenamefont {Couder}, \citenamefont {Fort},
  \citenamefont {Gautier},\ and\ \citenamefont {Boudaoud}}]{Couder2005}%
  \BibitemOpen
  \bibfield  {author} {\bibinfo {author} {\bibfnamefont {Y.}~\bibnamefont
  {Couder}}, \bibinfo {author} {\bibfnamefont {E.}~\bibnamefont {Fort}},
  \bibinfo {author} {\bibfnamefont {C.-H.}\ \bibnamefont {Gautier}},\ and\
  \bibinfo {author} {\bibfnamefont {A.}~\bibnamefont {Boudaoud}},\ }\bibfield
  {title} {\bibinfo {title} {From bouncing to floating: noncoalescence of drops
  on a fluid bath},\ }\href@noop {} {\bibfield  {journal} {\bibinfo  {journal}
  {Phys. Rev. Lett.}\ }\textbf {\bibinfo {volume} {94}},\ \bibinfo {pages}
  {177801} (\bibinfo {year} {2005}{\natexlab{a}})}\BibitemShut {NoStop}%
\bibitem [{\citenamefont {Couder}\ \emph
  {et~al.}(2005{\natexlab{b}})\citenamefont {Couder}, \citenamefont
  {Proti{\`{e}}re}, \citenamefont {Fort},\ and\ \citenamefont
  {Boudaoud}}]{Couder2005WalkingDroplets}%
  \BibitemOpen
  \bibfield  {author} {\bibinfo {author} {\bibfnamefont {Y.}~\bibnamefont
  {Couder}}, \bibinfo {author} {\bibfnamefont {S.}~\bibnamefont
  {Proti{\`{e}}re}}, \bibinfo {author} {\bibfnamefont {E.}~\bibnamefont
  {Fort}},\ and\ \bibinfo {author} {\bibfnamefont {A.}~\bibnamefont
  {Boudaoud}},\ }\bibfield  {title} {\bibinfo {title} {Dynamical phenomena:
  Walking and orbiting droplets},\ }\href@noop {} {\bibfield  {journal}
  {\bibinfo  {journal} {Nature}\ }\textbf {\bibinfo {volume} {437}},\ \bibinfo
  {pages} {208} (\bibinfo {year} {2005}{\natexlab{b}})}\BibitemShut {NoStop}%
\bibitem [{\citenamefont {Mol{\'a}{\v{c}}ek}\ and\ \citenamefont
  {Bush}(2013)}]{Molacek2013DropsTheory}%
  \BibitemOpen
  \bibfield  {author} {\bibinfo {author} {\bibfnamefont {J.}~\bibnamefont
  {Mol{\'a}{\v{c}}ek}}\ and\ \bibinfo {author} {\bibfnamefont {J.~W.~M.}\
  \bibnamefont {Bush}},\ }\bibfield  {title} {\bibinfo {title} {Drops walking
  on a vibrating bath: towards a hydrodynamic pilot-wave theory},\ }\href@noop
  {} {\bibfield  {journal} {\bibinfo  {journal} {J. Fluid Mech.}\ }\textbf
  {\bibinfo {volume} {727}},\ \bibinfo {pages} {612} (\bibinfo {year}
  {2013})}\BibitemShut {NoStop}%
\bibitem [{\citenamefont {Valani}\ \emph {et~al.}(2019)\citenamefont {Valani},
  \citenamefont {Slim},\ and\ \citenamefont {Simula}}]{superwalker}%
  \BibitemOpen
  \bibfield  {author} {\bibinfo {author} {\bibfnamefont {R.~N.}\ \bibnamefont
  {Valani}}, \bibinfo {author} {\bibfnamefont {A.~C.}\ \bibnamefont {Slim}},\
  and\ \bibinfo {author} {\bibfnamefont {T.}~\bibnamefont {Simula}},\
  }\bibfield  {title} {\bibinfo {title} {Superwalking droplets},\ }\href@noop
  {} {\bibfield  {journal} {\bibinfo  {journal} {Phys. Rev. Lett.}\ }\textbf
  {\bibinfo {volume} {123}},\ \bibinfo {pages} {024503} (\bibinfo {year}
  {2019})}\BibitemShut {NoStop}%
\bibitem [{\citenamefont {{Faraday}}(1831)}]{Faraday1831a}%
  \BibitemOpen
  \bibfield  {author} {\bibinfo {author} {\bibfnamefont {M.}~\bibnamefont
  {{Faraday}}},\ }\bibfield  {title} {\bibinfo {title} {{On a Peculiar Class of
  Acoustical Figures; and on Certain Forms Assumed by Groups of Particles upon
  Vibrating Elastic Surfaces}},\ }\href@noop {} {\bibfield  {journal} {\bibinfo
   {journal} {Phil. Trans. Roy. Soc. London Series I}\ }\textbf {\bibinfo
  {volume} {121}},\ \bibinfo {pages} {299} (\bibinfo {year}
  {1831})}\BibitemShut {NoStop}%
\bibitem [{\citenamefont {Fort}\ \emph {et~al.}(2010)\citenamefont {Fort},
  \citenamefont {Eddi}, \citenamefont {Boudaoud}, \citenamefont {Moukhtar},\
  and\ \citenamefont {Couder}}]{Fort17515}%
  \BibitemOpen
  \bibfield  {author} {\bibinfo {author} {\bibfnamefont {E.}~\bibnamefont
  {Fort}}, \bibinfo {author} {\bibfnamefont {A.}~\bibnamefont {Eddi}}, \bibinfo
  {author} {\bibfnamefont {A.}~\bibnamefont {Boudaoud}}, \bibinfo {author}
  {\bibfnamefont {J.}~\bibnamefont {Moukhtar}},\ and\ \bibinfo {author}
  {\bibfnamefont {Y.}~\bibnamefont {Couder}},\ }\bibfield  {title} {\bibinfo
  {title} {Path-memory induced quantization of classical orbits},\ }\href@noop
  {} {\bibfield  {journal} {\bibinfo  {journal} {Proc. Natl. Acad. Sci.}\
  }\textbf {\bibinfo {volume} {107}},\ \bibinfo {pages} {17515} (\bibinfo
  {year} {2010})}\BibitemShut {NoStop}%
\bibitem [{\citenamefont {Harris}\ and\ \citenamefont
  {Bush}(2014)}]{harris_bush_2014}%
  \BibitemOpen
  \bibfield  {author} {\bibinfo {author} {\bibfnamefont {D.~M.}\ \bibnamefont
  {Harris}}\ and\ \bibinfo {author} {\bibfnamefont {J.~W.~M.}\ \bibnamefont
  {Bush}},\ }\bibfield  {title} {\bibinfo {title} {Droplets walking in a
  rotating frame: from quantized orbits to multimodal statistics},\ }\href
  {https://doi.org/10.1017/jfm.2013.627} {\bibfield  {journal} {\bibinfo
  {journal} {J. Fluid Mech.}\ }\textbf {\bibinfo {volume} {739}},\ \bibinfo
  {pages} {444–464} (\bibinfo {year} {2014})}\BibitemShut {NoStop}%
\bibitem [{\citenamefont {Oza}\ \emph {et~al.}(2014)\citenamefont {Oza},
  \citenamefont {Harris}, \citenamefont {Rosales},\ and\ \citenamefont
  {Bush}}]{Oza2014}%
  \BibitemOpen
  \bibfield  {author} {\bibinfo {author} {\bibfnamefont {A.~U.}\ \bibnamefont
  {Oza}}, \bibinfo {author} {\bibfnamefont {D.~M.}\ \bibnamefont {Harris}},
  \bibinfo {author} {\bibfnamefont {R.~R.}\ \bibnamefont {Rosales}},\ and\
  \bibinfo {author} {\bibfnamefont {J.~W.~M.}\ \bibnamefont {Bush}},\
  }\bibfield  {title} {\bibinfo {title} {Pilot-wave dynamics in a rotating
  frame: on the emergence of orbital quantization},\ }\href@noop {} {\bibfield
  {journal} {\bibinfo  {journal} {J. Fluid Mech.}\ }\textbf {\bibinfo {volume}
  {744}},\ \bibinfo {pages} {404} (\bibinfo {year} {2014})}\BibitemShut
  {NoStop}%
\bibitem [{\citenamefont {Perrard}\ \emph
  {et~al.}(2014{\natexlab{a}})\citenamefont {Perrard}, \citenamefont
  {Labousse}, \citenamefont {Fort},\ and\ \citenamefont
  {Couder}}]{Perrard2014b}%
  \BibitemOpen
  \bibfield  {author} {\bibinfo {author} {\bibfnamefont {S.}~\bibnamefont
  {Perrard}}, \bibinfo {author} {\bibfnamefont {M.}~\bibnamefont {Labousse}},
  \bibinfo {author} {\bibfnamefont {E.}~\bibnamefont {Fort}},\ and\ \bibinfo
  {author} {\bibfnamefont {Y.}~\bibnamefont {Couder}},\ }\bibfield  {title}
  {\bibinfo {title} {Chaos driven by interfering memory},\ }\href@noop {}
  {\bibfield  {journal} {\bibinfo  {journal} {Phys. Rev. Lett.}\ }\textbf
  {\bibinfo {volume} {113}},\ \bibinfo {pages} {104101} (\bibinfo {year}
  {2014}{\natexlab{a}})}\BibitemShut {NoStop}%
\bibitem [{\citenamefont {Perrard}\ \emph
  {et~al.}(2014{\natexlab{b}})\citenamefont {Perrard}, \citenamefont
  {Labousse}, \citenamefont {Miskin}, \citenamefont {Fort},\ and\ \citenamefont
  {Couder}}]{Perrard2014a}%
  \BibitemOpen
  \bibfield  {author} {\bibinfo {author} {\bibfnamefont {S.}~\bibnamefont
  {Perrard}}, \bibinfo {author} {\bibfnamefont {M.}~\bibnamefont {Labousse}},
  \bibinfo {author} {\bibfnamefont {M.}~\bibnamefont {Miskin}}, \bibinfo
  {author} {\bibfnamefont {E.}~\bibnamefont {Fort}},\ and\ \bibinfo {author}
  {\bibfnamefont {Y.}~\bibnamefont {Couder}},\ }\bibfield  {title} {\bibinfo
  {title} {Self-organization into quantized eigenstates of a classical
  wave-driven particle},\ }\href@noop {} {\bibfield  {journal} {\bibinfo
  {journal} {Nat. Commun.}\ }\textbf {\bibinfo {volume} {5}},\ \bibinfo {pages}
  {3219} (\bibinfo {year} {2014}{\natexlab{b}})}\BibitemShut {NoStop}%
\bibitem [{\citenamefont {Labousse}\ \emph {et~al.}(2016)\citenamefont
  {Labousse}, \citenamefont {Perrard}, \citenamefont {Couder},\ and\
  \citenamefont {Fort}}]{labousse2016}%
  \BibitemOpen
  \bibfield  {author} {\bibinfo {author} {\bibfnamefont {M.}~\bibnamefont
  {Labousse}}, \bibinfo {author} {\bibfnamefont {S.}~\bibnamefont {Perrard}},
  \bibinfo {author} {\bibfnamefont {Y.}~\bibnamefont {Couder}},\ and\ \bibinfo
  {author} {\bibfnamefont {E.}~\bibnamefont {Fort}},\ }\bibfield  {title}
  {\bibinfo {title} {Self-attraction into spinning eigenstates of a mobile wave
  source by its emission back-reaction},\ }\href@noop {} {\bibfield  {journal}
  {\bibinfo  {journal} {Phys. Rev. E}\ }\textbf {\bibinfo {volume} {94}},\
  \bibinfo {pages} {042224} (\bibinfo {year} {2016})}\BibitemShut {NoStop}%
\bibitem [{\citenamefont {Eddi}\ \emph {et~al.}(2012)\citenamefont {Eddi},
  \citenamefont {Moukhtar}, \citenamefont {Perrard}, \citenamefont {Fort},\
  and\ \citenamefont {Couder}}]{Zeeman}%
  \BibitemOpen
  \bibfield  {author} {\bibinfo {author} {\bibfnamefont {A.}~\bibnamefont
  {Eddi}}, \bibinfo {author} {\bibfnamefont {J.}~\bibnamefont {Moukhtar}},
  \bibinfo {author} {\bibfnamefont {S.}~\bibnamefont {Perrard}}, \bibinfo
  {author} {\bibfnamefont {E.}~\bibnamefont {Fort}},\ and\ \bibinfo {author}
  {\bibfnamefont {Y.}~\bibnamefont {Couder}},\ }\bibfield  {title} {\bibinfo
  {title} {Level splitting at macroscopic scale},\ }\href
  {https://doi.org/10.1103/PhysRevLett.108.264503} {\bibfield  {journal}
  {\bibinfo  {journal} {Phys. Rev. Lett.}\ }\textbf {\bibinfo {volume} {108}},\
  \bibinfo {pages} {264503} (\bibinfo {year} {2012})}\BibitemShut {NoStop}%
\bibitem [{\citenamefont {Oza}\ \emph {et~al.}(2018{\natexlab{a}})\citenamefont
  {Oza}, \citenamefont {Rosales},\ and\ \citenamefont {Bush}}]{spinstates2018}%
  \BibitemOpen
  \bibfield  {author} {\bibinfo {author} {\bibfnamefont {A.~U.}\ \bibnamefont
  {Oza}}, \bibinfo {author} {\bibfnamefont {R.~R.}\ \bibnamefont {Rosales}},\
  and\ \bibinfo {author} {\bibfnamefont {J.~W.~M.}\ \bibnamefont {Bush}},\
  }\bibfield  {title} {\bibinfo {title} {Hydrodynamic spin states},\ }\href
  {https://doi.org/10.1063/1.5034134} {\bibfield  {journal} {\bibinfo
  {journal} {Chaos}\ }\textbf {\bibinfo {volume} {28}},\ \bibinfo {pages}
  {096106} (\bibinfo {year} {2018}{\natexlab{a}})}\BibitemShut {NoStop}%
\bibitem [{\citenamefont {Harris}\ \emph {et~al.}(2013)\citenamefont {Harris},
  \citenamefont {Moukhtar}, \citenamefont {Fort}, \citenamefont {Couder},\ and\
  \citenamefont {Bush}}]{PhysRevE.88.011001}%
  \BibitemOpen
  \bibfield  {author} {\bibinfo {author} {\bibfnamefont {D.~M.}\ \bibnamefont
  {Harris}}, \bibinfo {author} {\bibfnamefont {J.}~\bibnamefont {Moukhtar}},
  \bibinfo {author} {\bibfnamefont {E.}~\bibnamefont {Fort}}, \bibinfo {author}
  {\bibfnamefont {Y.}~\bibnamefont {Couder}},\ and\ \bibinfo {author}
  {\bibfnamefont {J.~W.~M.}\ \bibnamefont {Bush}},\ }\bibfield  {title}
  {\bibinfo {title} {Wavelike statistics from pilot-wave dynamics in a circular
  corral},\ }\href@noop {} {\bibfield  {journal} {\bibinfo  {journal} {Phys.
  Rev. E}\ }\textbf {\bibinfo {volume} {88}},\ \bibinfo {pages} {011001}
  (\bibinfo {year} {2013})}\BibitemShut {NoStop}%
\bibitem [{\citenamefont {Gilet}(2016)}]{Giletconfined2016}%
  \BibitemOpen
  \bibfield  {author} {\bibinfo {author} {\bibfnamefont {T.}~\bibnamefont
  {Gilet}},\ }\bibfield  {title} {\bibinfo {title} {Quantumlike statistics of
  deterministic wave-particle interactions in a circular cavity},\ }\href
  {https://doi.org/10.1103/PhysRevE.93.042202} {\bibfield  {journal} {\bibinfo
  {journal} {Phys. Rev. E}\ }\textbf {\bibinfo {volume} {93}},\ \bibinfo
  {pages} {042202} (\bibinfo {year} {2016})}\BibitemShut {NoStop}%
\bibitem [{\citenamefont {S{\'a}enz}\ \emph {et~al.}(2018)\citenamefont
  {S{\'a}enz}, \citenamefont {Cristea-Platon},\ and\ \citenamefont
  {Bush}}]{Saenz2017}%
  \BibitemOpen
  \bibfield  {author} {\bibinfo {author} {\bibfnamefont {P.~J.}\ \bibnamefont
  {S{\'a}enz}}, \bibinfo {author} {\bibfnamefont {T.}~\bibnamefont
  {Cristea-Platon}},\ and\ \bibinfo {author} {\bibfnamefont {J.~W.~M.}\
  \bibnamefont {Bush}},\ }\bibfield  {title} {\bibinfo {title} {Statistical
  projection effects in a hydrodynamic pilot-wave system},\ }\href@noop {}
  {\bibfield  {journal} {\bibinfo  {journal} {Nat. Phys.}\ }\textbf {\bibinfo
  {volume} {14}},\ \bibinfo {pages} {315} (\bibinfo {year} {2018})}\BibitemShut
  {NoStop}%
\bibitem [{\citenamefont {Cristea-Platon}\ \emph {et~al.}(2018)\citenamefont
  {Cristea-Platon}, \citenamefont {S\'{a}enz},\ and\ \citenamefont
  {Bush}}]{Cristea}%
  \BibitemOpen
  \bibfield  {author} {\bibinfo {author} {\bibfnamefont {T.}~\bibnamefont
  {Cristea-Platon}}, \bibinfo {author} {\bibfnamefont {P.~J.}\ \bibnamefont
  {S\'{a}enz}},\ and\ \bibinfo {author} {\bibfnamefont {J.~W.~M.}\ \bibnamefont
  {Bush}},\ }\bibfield  {title} {\bibinfo {title} {Walking droplets in a
  circular corral: Quantisation and chaos},\ }\href@noop {} {\bibfield
  {journal} {\bibinfo  {journal} {Chaos}\ }\textbf {\bibinfo {volume} {28}},\
  \bibinfo {pages} {096116} (\bibinfo {year} {2018})}\BibitemShut {NoStop}%
\bibitem [{\citenamefont {Durey}\ \emph
  {et~al.}(2020{\natexlab{a}})\citenamefont {Durey}, \citenamefont {Milewski},\
  and\ \citenamefont {Wang}}]{durey_milewski_wang_2020}%
  \BibitemOpen
  \bibfield  {author} {\bibinfo {author} {\bibfnamefont {M.}~\bibnamefont
  {Durey}}, \bibinfo {author} {\bibfnamefont {P.~A.}\ \bibnamefont
  {Milewski}},\ and\ \bibinfo {author} {\bibfnamefont {Z.}~\bibnamefont
  {Wang}},\ }\bibfield  {title} {\bibinfo {title} {Faraday pilot-wave dynamics
  in a circular corral},\ }\href@noop {} {\bibfield  {journal} {\bibinfo
  {journal} {J. Fluid Mech.}\ }\textbf {\bibinfo {volume} {891}},\ \bibinfo
  {pages} {A3} (\bibinfo {year} {2020}{\natexlab{a}})}\BibitemShut {NoStop}%
\bibitem [{\citenamefont {S{\'a}enz}\ \emph {et~al.}(2020)\citenamefont
  {S{\'a}enz}, \citenamefont {Cristea-Platon},\ and\ \citenamefont
  {Bush}}]{Friedal}%
  \BibitemOpen
  \bibfield  {author} {\bibinfo {author} {\bibfnamefont {P.~J.}\ \bibnamefont
  {S{\'a}enz}}, \bibinfo {author} {\bibfnamefont {T.}~\bibnamefont
  {Cristea-Platon}},\ and\ \bibinfo {author} {\bibfnamefont {J.~W.~M.}\
  \bibnamefont {Bush}},\ }\bibfield  {title} {\bibinfo {title} {A hydrodynamic
  analog of {F}riedel oscillations},\ }\href@noop {} {\bibfield  {journal}
  {\bibinfo  {journal} {Sci. Adv.}\ }\textbf {\bibinfo {volume} {6}} (\bibinfo
  {year} {2020})}\BibitemShut {NoStop}%
\bibitem [{\citenamefont {Eddi}\ \emph {et~al.}(2009)\citenamefont {Eddi},
  \citenamefont {Fort}, \citenamefont {Moisy},\ and\ \citenamefont
  {Couder}}]{Eddi2009}%
  \BibitemOpen
  \bibfield  {author} {\bibinfo {author} {\bibfnamefont {A.}~\bibnamefont
  {Eddi}}, \bibinfo {author} {\bibfnamefont {E.}~\bibnamefont {Fort}}, \bibinfo
  {author} {\bibfnamefont {F.}~\bibnamefont {Moisy}},\ and\ \bibinfo {author}
  {\bibfnamefont {Y.}~\bibnamefont {Couder}},\ }\bibfield  {title} {\bibinfo
  {title} {Unpredictable tunneling of a classical wave-particle association},\
  }\href@noop {} {\bibfield  {journal} {\bibinfo  {journal} {Phys. Rev. Lett.}\
  }\textbf {\bibinfo {volume} {102}},\ \bibinfo {pages} {240401} (\bibinfo
  {year} {2009})}\BibitemShut {NoStop}%
\bibitem [{\citenamefont {Nachbin}\ \emph {et~al.}(2017)\citenamefont
  {Nachbin}, \citenamefont {Milewski},\ and\ \citenamefont
  {Bush}}]{tunnelingnachbin}%
  \BibitemOpen
  \bibfield  {author} {\bibinfo {author} {\bibfnamefont {A.}~\bibnamefont
  {Nachbin}}, \bibinfo {author} {\bibfnamefont {P.~A.}\ \bibnamefont
  {Milewski}},\ and\ \bibinfo {author} {\bibfnamefont {J.~W.~M.}\ \bibnamefont
  {Bush}},\ }\bibfield  {title} {\bibinfo {title} {Tunneling with a
  hydrodynamic pilot-wave model},\ }\href
  {https://doi.org/10.1103/PhysRevFluids.2.034801} {\bibfield  {journal}
  {\bibinfo  {journal} {Phys. Rev. Fluids}\ }\textbf {\bibinfo {volume} {2}},\
  \bibinfo {pages} {034801} (\bibinfo {year} {2017})}\BibitemShut {NoStop}%
\bibitem [{\citenamefont {Tadrist}\ \emph {et~al.}(2020)\citenamefont
  {Tadrist}, \citenamefont {Gilet}, \citenamefont {Schlagheck},\ and\
  \citenamefont {Bush}}]{tunneling2020}%
  \BibitemOpen
  \bibfield  {author} {\bibinfo {author} {\bibfnamefont {L.}~\bibnamefont
  {Tadrist}}, \bibinfo {author} {\bibfnamefont {T.}~\bibnamefont {Gilet}},
  \bibinfo {author} {\bibfnamefont {P.}~\bibnamefont {Schlagheck}},\ and\
  \bibinfo {author} {\bibfnamefont {J.~W.~M.}\ \bibnamefont {Bush}},\
  }\bibfield  {title} {\bibinfo {title} {Predictability in a hydrodynamic
  pilot-wave system: Resolution of walker tunneling},\ }\href
  {https://doi.org/10.1103/PhysRevE.102.013104} {\bibfield  {journal} {\bibinfo
   {journal} {Phys. Rev. E}\ }\textbf {\bibinfo {volume} {102}},\ \bibinfo
  {pages} {013104} (\bibinfo {year} {2020})}\BibitemShut {NoStop}%
\bibitem [{\citenamefont {Valani}\ \emph {et~al.}(2018)\citenamefont {Valani},
  \citenamefont {Slim},\ and\ \citenamefont {Simula}}]{ValaniHOM}%
  \BibitemOpen
  \bibfield  {author} {\bibinfo {author} {\bibfnamefont {R.~N.}\ \bibnamefont
  {Valani}}, \bibinfo {author} {\bibfnamefont {A.~C.}\ \bibnamefont {Slim}},\
  and\ \bibinfo {author} {\bibfnamefont {T.}~\bibnamefont {Simula}},\
  }\bibfield  {title} {\bibinfo {title} {Hong–{O}u–{M}andel-like
  two-droplet correlations},\ }\href@noop {} {\bibfield  {journal} {\bibinfo
  {journal} {Chaos}\ }\textbf {\bibinfo {volume} {28}},\ \bibinfo {pages}
  {096104} (\bibinfo {year} {2018})}\BibitemShut {NoStop}%
\bibitem [{\citenamefont {Nachbin}(2018)}]{correlationnachbin}%
  \BibitemOpen
  \bibfield  {author} {\bibinfo {author} {\bibfnamefont {A.}~\bibnamefont
  {Nachbin}},\ }\bibfield  {title} {\bibinfo {title} {Walking droplets
  correlated at a distance},\ }\href {https://doi.org/10.1063/1.5050805}
  {\bibfield  {journal} {\bibinfo  {journal} {Chaos}\ }\textbf {\bibinfo
  {volume} {28}},\ \bibinfo {pages} {096110} (\bibinfo {year}
  {2018})}\BibitemShut {NoStop}%
\bibitem [{\citenamefont {Dagan}\ and\ \citenamefont
  {Bush}(2020)}]{Dagan2020hqft}%
  \BibitemOpen
  \bibfield  {author} {\bibinfo {author} {\bibfnamefont {Y.}~\bibnamefont
  {Dagan}}\ and\ \bibinfo {author} {\bibfnamefont {J.~W.~M.}\ \bibnamefont
  {Bush}},\ }\bibfield  {title} {\bibinfo {title} {Hydrodynamic quantum field
  theory: the free particle},\ }\href {https://doi.org/10.5802/crmeca.34}
  {\bibfield  {journal} {\bibinfo  {journal} {Comptes Rendus. M{\'e}canique}\
  }\textbf {\bibinfo {volume} {348}},\ \bibinfo {pages} {555} (\bibinfo {year}
  {2020})}\BibitemShut {NoStop}%
\bibitem [{\citenamefont {Durey}\ and\ \citenamefont
  {Bush}(2020)}]{Durey2020hqft}%
  \BibitemOpen
  \bibfield  {author} {\bibinfo {author} {\bibfnamefont {M.}~\bibnamefont
  {Durey}}\ and\ \bibinfo {author} {\bibfnamefont {J.~W.~M.}\ \bibnamefont
  {Bush}},\ }\bibfield  {title} {\bibinfo {title} {Hydrodynamic quantum field
  theory: The onset of particle motion and the form of the pilot wave},\ }\href
  {https://doi.org/10.3389/fphy.2020.00300} {\bibfield  {journal} {\bibinfo
  {journal} {Front. Phys.}\ }\textbf {\bibinfo {volume} {8}},\ \bibinfo {pages}
  {300} (\bibinfo {year} {2020})}\BibitemShut {NoStop}%
\bibitem [{\citenamefont {Bush}(2015)}]{Bush2015}%
  \BibitemOpen
  \bibfield  {author} {\bibinfo {author} {\bibfnamefont {J.~W.~M.}\
  \bibnamefont {Bush}},\ }\bibfield  {title} {\bibinfo {title} {Pilot-wave
  hydrodynamics},\ }\href@noop {} {\bibfield  {journal} {\bibinfo  {journal}
  {Annu. Rev. Fluid Mech.}\ }\textbf {\bibinfo {volume} {47}},\ \bibinfo
  {pages} {269} (\bibinfo {year} {2015})}\BibitemShut {NoStop}%
\bibitem [{\citenamefont {Bush}\ and\ \citenamefont {Oza}(2020)}]{Bush_2020}%
  \BibitemOpen
  \bibfield  {author} {\bibinfo {author} {\bibfnamefont {J.~W.~M.}\
  \bibnamefont {Bush}}\ and\ \bibinfo {author} {\bibfnamefont {A.~U.}\
  \bibnamefont {Oza}},\ }\bibfield  {title} {\bibinfo {title} {Hydrodynamic
  quantum analogs},\ }\href {https://doi.org/10.1088/1361-6633/abc22c}
  {\bibfield  {journal} {\bibinfo  {journal} {Rep. Prog. Phys.}\ }\textbf
  {\bibinfo {volume} {84}},\ \bibinfo {pages} {017001} (\bibinfo {year}
  {2020})}\BibitemShut {NoStop}%
\bibitem [{\citenamefont {Turton}\ \emph {et~al.}(2018)\citenamefont {Turton},
  \citenamefont {Couchman},\ and\ \citenamefont {Bush}}]{Turton2018}%
  \BibitemOpen
  \bibfield  {author} {\bibinfo {author} {\bibfnamefont {S.~E.}\ \bibnamefont
  {Turton}}, \bibinfo {author} {\bibfnamefont {M.~M.~P.}\ \bibnamefont
  {Couchman}},\ and\ \bibinfo {author} {\bibfnamefont {J.~W.~M.}\ \bibnamefont
  {Bush}},\ }\bibfield  {title} {\bibinfo {title} {A review of the theoretical
  modeling of walking droplets: Toward a generalized pilot-wave framework},\
  }\href@noop {} {\bibfield  {journal} {\bibinfo  {journal} {Chaos}\ }\textbf
  {\bibinfo {volume} {28}},\ \bibinfo {pages} {096111} (\bibinfo {year}
  {2018})}\BibitemShut {NoStop}%
\bibitem [{\citenamefont {Rahman}\ and\ \citenamefont
  {Blackmore}(2020)}]{Rahman2020review}%
  \BibitemOpen
  \bibfield  {author} {\bibinfo {author} {\bibfnamefont {A.}~\bibnamefont
  {Rahman}}\ and\ \bibinfo {author} {\bibfnamefont {D.}~\bibnamefont
  {Blackmore}},\ }\bibfield  {title} {\bibinfo {title} {Walking droplets
  through the lens of dynamical systems},\ }\href
  {https://doi.org/10.1142/S0217984920300094} {\bibfield  {journal} {\bibinfo
  {journal} {Mod. Phys. Lett. B}\ }\textbf {\bibinfo {volume} {34}},\ \bibinfo
  {pages} {2030009} (\bibinfo {year} {2020})}\BibitemShut {NoStop}%
\bibitem [{\citenamefont {Valani}\ \emph {et~al.}(2021)\citenamefont {Valani},
  \citenamefont {Dring}, \citenamefont {Simula},\ and\ \citenamefont
  {Slim}}]{superwalkernumerical}%
  \BibitemOpen
  \bibfield  {author} {\bibinfo {author} {\bibfnamefont {R.~N.}\ \bibnamefont
  {Valani}}, \bibinfo {author} {\bibfnamefont {J.}~\bibnamefont {Dring}},
  \bibinfo {author} {\bibfnamefont {T.~P.}\ \bibnamefont {Simula}},\ and\
  \bibinfo {author} {\bibfnamefont {A.~C.}\ \bibnamefont {Slim}},\ }\bibfield
  {title} {\bibinfo {title} {Emergence of superwalking droplets},\ }\href
  {https://doi.org/10.1017/jfm.2020.742} {\bibfield  {journal} {\bibinfo
  {journal} {J. Fluid Mech.}\ }\textbf {\bibinfo {volume} {906}},\ \bibinfo
  {pages} {A3} (\bibinfo {year} {2021})}\BibitemShut {NoStop}%
\bibitem [{\citenamefont {Bacot}\ \emph {et~al.}(2019)\citenamefont {Bacot},
  \citenamefont {Perrard}, \citenamefont {Labousse}, \citenamefont {Couder},\
  and\ \citenamefont {Fort}}]{Bacot2019}%
  \BibitemOpen
  \bibfield  {author} {\bibinfo {author} {\bibfnamefont {V.}~\bibnamefont
  {Bacot}}, \bibinfo {author} {\bibfnamefont {S.}~\bibnamefont {Perrard}},
  \bibinfo {author} {\bibfnamefont {M.}~\bibnamefont {Labousse}}, \bibinfo
  {author} {\bibfnamefont {Y.}~\bibnamefont {Couder}},\ and\ \bibinfo {author}
  {\bibfnamefont {E.}~\bibnamefont {Fort}},\ }\bibfield  {title} {\bibinfo
  {title} {Multistable free states of an active particle from a coherent memory
  dynamics},\ }\href {https://doi.org/10.1103/PhysRevLett.122.104303}
  {\bibfield  {journal} {\bibinfo  {journal} {Phys. Rev. Lett.}\ }\textbf
  {\bibinfo {volume} {122}},\ \bibinfo {pages} {104303} (\bibinfo {year}
  {2019})}\BibitemShut {NoStop}%
\bibitem [{\citenamefont {Hubert}\ \emph {et~al.}(2019)\citenamefont {Hubert},
  \citenamefont {Perrard}, \citenamefont {Labousse}, \citenamefont
  {Vandewalle},\ and\ \citenamefont {Couder}}]{Hubert2019}%
  \BibitemOpen
  \bibfield  {author} {\bibinfo {author} {\bibfnamefont {M.}~\bibnamefont
  {Hubert}}, \bibinfo {author} {\bibfnamefont {S.}~\bibnamefont {Perrard}},
  \bibinfo {author} {\bibfnamefont {M.}~\bibnamefont {Labousse}}, \bibinfo
  {author} {\bibfnamefont {N.}~\bibnamefont {Vandewalle}},\ and\ \bibinfo
  {author} {\bibfnamefont {Y.}~\bibnamefont {Couder}},\ }\bibfield  {title}
  {\bibinfo {title} {Tunable bimodal explorations of space from memory-driven
  deterministic dynamics},\ }\href@noop {} {\bibfield  {journal} {\bibinfo
  {journal} {Phys. Rev. E}\ }\textbf {\bibinfo {volume} {100}},\ \bibinfo
  {pages} {032201} (\bibinfo {year} {2019})}\BibitemShut {NoStop}%
\bibitem [{\citenamefont {Berg}\ and\ \citenamefont {Brown}(1972)}]{BERG1972}%
  \BibitemOpen
  \bibfield  {author} {\bibinfo {author} {\bibfnamefont {H.~C.}\ \bibnamefont
  {Berg}}\ and\ \bibinfo {author} {\bibfnamefont {D.~A.}\ \bibnamefont
  {Brown}},\ }\bibfield  {title} {\bibinfo {title} {Chemotaxis in {E}scherichia
  coli analysed by three-dimensional tracking},\ }\href@noop {} {\bibfield
  {journal} {\bibinfo  {journal} {Nature}\ }\textbf {\bibinfo {volume} {239}},\
  \bibinfo {pages} {500} (\bibinfo {year} {1972})}\BibitemShut {NoStop}%
\bibitem [{\citenamefont {Hokmabad}\ \emph {et~al.}(2021)\citenamefont
  {Hokmabad}, \citenamefont {Dey}, \citenamefont {Jalaal}, \citenamefont
  {Mohanty}, \citenamefont {Almukambetova}, \citenamefont {Baldwin},
  \citenamefont {Lohse},\ and\ \citenamefont {Maass}}]{stopgoswim}%
  \BibitemOpen
  \bibfield  {author} {\bibinfo {author} {\bibfnamefont {B.~V.}\ \bibnamefont
  {Hokmabad}}, \bibinfo {author} {\bibfnamefont {R.}~\bibnamefont {Dey}},
  \bibinfo {author} {\bibfnamefont {M.}~\bibnamefont {Jalaal}}, \bibinfo
  {author} {\bibfnamefont {D.}~\bibnamefont {Mohanty}}, \bibinfo {author}
  {\bibfnamefont {M.}~\bibnamefont {Almukambetova}}, \bibinfo {author}
  {\bibfnamefont {K.~A.}\ \bibnamefont {Baldwin}}, \bibinfo {author}
  {\bibfnamefont {D.}~\bibnamefont {Lohse}},\ and\ \bibinfo {author}
  {\bibfnamefont {C.~C.}\ \bibnamefont {Maass}},\ }\bibfield  {title} {\bibinfo
  {title} {Emergence of bimodal motility in active droplets},\ }\href
  {https://doi.org/10.1103/PhysRevX.11.011043} {\bibfield  {journal} {\bibinfo
  {journal} {Phys. Rev. X}\ }\textbf {\bibinfo {volume} {11}},\ \bibinfo
  {pages} {011043} (\bibinfo {year} {2021})}\BibitemShut {NoStop}%
\bibitem [{\citenamefont {Stocker}(2011)}]{Stocker2635}%
  \BibitemOpen
  \bibfield  {author} {\bibinfo {author} {\bibfnamefont {R.}~\bibnamefont
  {Stocker}},\ }\bibfield  {title} {\bibinfo {title} {Reverse and flick: Hybrid
  locomotion in bacteria},\ }\href@noop {} {\bibfield  {journal} {\bibinfo
  {journal} {Proc. Natl. Acad. Sci.}\ }\textbf {\bibinfo {volume} {108}},\
  \bibinfo {pages} {2635} (\bibinfo {year} {2011})}\BibitemShut {NoStop}%
\bibitem [{\citenamefont {Bhattacharjee}\ and\ \citenamefont
  {Datta}(2019)}]{Bhattacharjee2019}%
  \BibitemOpen
  \bibfield  {author} {\bibinfo {author} {\bibfnamefont {T.}~\bibnamefont
  {Bhattacharjee}}\ and\ \bibinfo {author} {\bibfnamefont {S.~S.}\ \bibnamefont
  {Datta}},\ }\bibfield  {title} {\bibinfo {title} {Bacterial hopping and
  trapping in porous media},\ }\href
  {https://doi.org/10.1038/s41467-019-10115-1} {\bibfield  {journal} {\bibinfo
  {journal} {Nat. Commun.}\ }\textbf {\bibinfo {volume} {10}},\ \bibinfo
  {pages} {2075} (\bibinfo {year} {2019})}\BibitemShut {NoStop}%
\bibitem [{\citenamefont {Durey}\ \emph
  {et~al.}(2020{\natexlab{b}})\citenamefont {Durey}, \citenamefont {Turton},\
  and\ \citenamefont {Bush}}]{Durey2020}%
  \BibitemOpen
  \bibfield  {author} {\bibinfo {author} {\bibfnamefont {M.}~\bibnamefont
  {Durey}}, \bibinfo {author} {\bibfnamefont {S.~E.}\ \bibnamefont {Turton}},\
  and\ \bibinfo {author} {\bibfnamefont {J.~W.~M.}\ \bibnamefont {Bush}},\
  }\bibfield  {title} {\bibinfo {title} {Speed oscillations in classical
  pilot-wave dynamics},\ }\href@noop {} {\bibfield  {journal} {\bibinfo
  {journal} {Proc. Math. Phys. Eng. Sci.}\ }\textbf {\bibinfo {volume} {476}},\
  \bibinfo {pages} {20190884} (\bibinfo {year}
  {2020}{\natexlab{b}})}\BibitemShut {NoStop}%
\bibitem [{\citenamefont {Oza}\ \emph {et~al.}(2013)\citenamefont {Oza},
  \citenamefont {Rosales},\ and\ \citenamefont {Bush}}]{Oza2013}%
  \BibitemOpen
  \bibfield  {author} {\bibinfo {author} {\bibfnamefont {A.~U.}\ \bibnamefont
  {Oza}}, \bibinfo {author} {\bibfnamefont {R.~R.}\ \bibnamefont {Rosales}},\
  and\ \bibinfo {author} {\bibfnamefont {J.~W.~M.}\ \bibnamefont {Bush}},\
  }\bibfield  {title} {\bibinfo {title} {A trajectory equation for walking
  droplets: hydrodynamic pilot-wave theory},\ }\href@noop {} {\bibfield
  {journal} {\bibinfo  {journal} {J. Fluid Mech.}\ }\textbf {\bibinfo {volume}
  {737}},\ \bibinfo {pages} {552} (\bibinfo {year} {2013})}\BibitemShut
  {NoStop}%
\bibitem [{\citenamefont {Valani}(2020)}]{phdthesisrahil}%
  \BibitemOpen
  \bibfield  {author} {\bibinfo {author} {\bibfnamefont {R.~N.}\ \bibnamefont
  {Valani}},\ }\emph {\bibinfo {title} {Superwalking Droplets and Generalised
  Pilot-Wave Dynamics}},\ \href@noop {} {Ph.D. thesis},\ \bibinfo  {school}
  {Monash University} (\bibinfo {year} {2020})\BibitemShut {NoStop}%
\bibitem [{\citenamefont {Durey}(2020)}]{Durey2020lorenz}%
  \BibitemOpen
  \bibfield  {author} {\bibinfo {author} {\bibfnamefont {M.}~\bibnamefont
  {Durey}},\ }\bibfield  {title} {\bibinfo {title} {Bifurcations and chaos in a
  {L}orenz-like pilot-wave system},\ }\href@noop {} {\bibfield  {journal}
  {\bibinfo  {journal} {Chaos}\ }\textbf {\bibinfo {volume} {30}},\ \bibinfo
  {pages} {103115} (\bibinfo {year} {2020})}\BibitemShut {NoStop}%
\bibitem [{\citenamefont {Durey}\ \emph {et~al.}(2018)\citenamefont {Durey},
  \citenamefont {Milewski},\ and\ \citenamefont {Bush}}]{durey2018}%
  \BibitemOpen
  \bibfield  {author} {\bibinfo {author} {\bibfnamefont {M.}~\bibnamefont
  {Durey}}, \bibinfo {author} {\bibfnamefont {P.~A.}\ \bibnamefont
  {Milewski}},\ and\ \bibinfo {author} {\bibfnamefont {J.~W.~M.}\ \bibnamefont
  {Bush}},\ }\bibfield  {title} {\bibinfo {title} {Dynamics, emergent
  statistics, and the mean-pilot-wave potential of walking droplets},\
  }\href@noop {} {\bibfield  {journal} {\bibinfo  {journal} {Chaos}\ }\textbf
  {\bibinfo {volume} {28}},\ \bibinfo {pages} {096108} (\bibinfo {year}
  {2018})}\BibitemShut {NoStop}%
\bibitem [{\citenamefont {Fort}\ and\ \citenamefont
  {Couder}(2013)}]{Fort_2013}%
  \BibitemOpen
  \bibfield  {author} {\bibinfo {author} {\bibfnamefont {E.}~\bibnamefont
  {Fort}}\ and\ \bibinfo {author} {\bibfnamefont {Y.}~\bibnamefont {Couder}},\
  }\bibfield  {title} {\bibinfo {title} {Trajectory eigenmodes of an orbiting
  wave source},\ }\href@noop {} {\bibfield  {journal} {\bibinfo  {journal}
  {Europhys. Lett.}\ }\textbf {\bibinfo {volume} {102}},\ \bibinfo {pages}
  {16005} (\bibinfo {year} {2013})}\BibitemShut {NoStop}%
\bibitem [{\citenamefont {Oza}\ \emph {et~al.}(2018{\natexlab{b}})\citenamefont
  {Oza}, \citenamefont {Rosales},\ and\ \citenamefont {Bush}}]{Spinstates}%
  \BibitemOpen
  \bibfield  {author} {\bibinfo {author} {\bibfnamefont {A.~U.}\ \bibnamefont
  {Oza}}, \bibinfo {author} {\bibfnamefont {R.~R.}\ \bibnamefont {Rosales}},\
  and\ \bibinfo {author} {\bibfnamefont {J.~W.~M.}\ \bibnamefont {Bush}},\
  }\bibfield  {title} {\bibinfo {title} {Hydrodynamic spin states},\
  }\href@noop {} {\bibfield  {journal} {\bibinfo  {journal} {Chaos}\ }\textbf
  {\bibinfo {volume} {28}},\ \bibinfo {pages} {096106} (\bibinfo {year}
  {2018}{\natexlab{b}})}\BibitemShut {NoStop}%
\bibitem [{\citenamefont {Beck}\ and\ \citenamefont
  {Sch\"{o}gl}(1993)}]{beck_schogl_1993}%
  \BibitemOpen
  \bibfield  {author} {\bibinfo {author} {\bibfnamefont {C.}~\bibnamefont
  {Beck}}\ and\ \bibinfo {author} {\bibfnamefont {F.}~\bibnamefont
  {Sch\"{o}gl}},\ }\href {https://doi.org/10.1017/CBO9780511524585} {\emph
  {\bibinfo {title} {Thermodynamics of Chaotic Systems: An Introduction}}},\
  Cambridge Nonlinear Science Series\ (\bibinfo  {publisher} {Cambridge
  University Press},\ \bibinfo {year} {1993})\BibitemShut {NoStop}%
\bibitem [{\citenamefont {Balakrishnan}(2020)}]{Balakrishnan2020}%
  \BibitemOpen
  \bibfield  {author} {\bibinfo {author} {\bibfnamefont {V.}~\bibnamefont
  {Balakrishnan}},\ }\href@noop {} {\emph {\bibinfo {title} {Mathematical
  Physics}}}\ (\bibinfo  {publisher} {Springer International Publishing},\
  \bibinfo {year} {2020})\BibitemShut {NoStop}%
\bibitem [{\citenamefont {Sparrow}(1982)}]{Sparrowbook}%
  \BibitemOpen
  \bibfield  {author} {\bibinfo {author} {\bibfnamefont {C.}~\bibnamefont
  {Sparrow}},\ }\href@noop {} {\emph {\bibinfo {title} {The Lorenz Equations:
  Bifurcations, Chaos, and Strange Attractors}}}\ (\bibinfo  {publisher}
  {Springer-Verlag, New York},\ \bibinfo {year} {1982})\BibitemShut {NoStop}%
\bibitem [{\citenamefont {Lorenz}(1963)}]{Lorenz1963}%
  \BibitemOpen
  \bibfield  {author} {\bibinfo {author} {\bibfnamefont {E.~N.}\ \bibnamefont
  {Lorenz}},\ }\bibfield  {title} {\bibinfo {title} {{Deterministic Nonperiodic
  Flow}},\ }\href@noop {} {\bibfield  {journal} {\bibinfo  {journal} {J Atmos
  Sci.}\ }\textbf {\bibinfo {volume} {20}},\ \bibinfo {pages} {130} (\bibinfo
  {year} {1963})}\BibitemShut {NoStop}%
\bibitem [{\citenamefont {Takeyama}(1978)}]{Takeyama1978_1}%
  \BibitemOpen
  \bibfield  {author} {\bibinfo {author} {\bibfnamefont {K.}~\bibnamefont
  {Takeyama}},\ }\bibfield  {title} {\bibinfo {title} {Dynamics of the {L}orenz
  model of convective instabilities},\ }\href@noop {} {\bibfield  {journal}
  {\bibinfo  {journal} {Prog. Theor. Phys.}\ }\textbf {\bibinfo {volume}
  {60}},\ \bibinfo {pages} {613} (\bibinfo {year} {1978})}\BibitemShut
  {NoStop}%
\bibitem [{\citenamefont {Festa}\ \emph
  {et~al.}(2002{\natexlab{a}})\citenamefont {Festa}, \citenamefont {Mazzino},\
  and\ \citenamefont {Vincenzi}}]{Festa_2002}%
  \BibitemOpen
  \bibfield  {author} {\bibinfo {author} {\bibfnamefont {R.}~\bibnamefont
  {Festa}}, \bibinfo {author} {\bibfnamefont {A.}~\bibnamefont {Mazzino}},\
  and\ \bibinfo {author} {\bibfnamefont {D.}~\bibnamefont {Vincenzi}},\
  }\bibfield  {title} {\bibinfo {title} {Lorenz deterministic diffusion},\
  }\href@noop {} {\bibfield  {journal} {\bibinfo  {journal} {Europhys. Lett.}\
  }\textbf {\bibinfo {volume} {60}},\ \bibinfo {pages} {820} (\bibinfo {year}
  {2002}{\natexlab{a}})}\BibitemShut {NoStop}%
\bibitem [{\citenamefont {Takeyama}(1980)}]{Takeyama1980_2}%
  \BibitemOpen
  \bibfield  {author} {\bibinfo {author} {\bibfnamefont {K.}~\bibnamefont
  {Takeyama}},\ }\bibfield  {title} {\bibinfo {title} {Dynamics of the {L}orenz
  model of convective instabilities. {II}},\ }\href@noop {} {\bibfield
  {journal} {\bibinfo  {journal} {Prog. Theor. Phys.}\ }\textbf {\bibinfo
  {volume} {63}},\ \bibinfo {pages} {91} (\bibinfo {year} {1980})}\BibitemShut
  {NoStop}%
\bibitem [{\citenamefont {Festa}\ \emph
  {et~al.}(2002{\natexlab{b}})\citenamefont {Festa}, \citenamefont {Mazzino},\
  and\ \citenamefont {Vincenzi}}]{Festa2002pre}%
  \BibitemOpen
  \bibfield  {author} {\bibinfo {author} {\bibfnamefont {R.}~\bibnamefont
  {Festa}}, \bibinfo {author} {\bibfnamefont {A.}~\bibnamefont {Mazzino}},\
  and\ \bibinfo {author} {\bibfnamefont {D.}~\bibnamefont {Vincenzi}},\
  }\bibfield  {title} {\bibinfo {title} {Lorenz-like systems and classical
  dynamical equations with memory forcing: An alternate point of view for
  singling out the origin of chaos},\ }\href@noop {} {\bibfield  {journal}
  {\bibinfo  {journal} {Phys. Rev. E}\ }\textbf {\bibinfo {volume} {65}},\
  \bibinfo {pages} {046205} (\bibinfo {year} {2002}{\natexlab{b}})}\BibitemShut
  {NoStop}%
\bibitem [{\citenamefont {Budanur}\ and\ \citenamefont
  {Fleury}(2019)}]{doi:10.1063/1.5058279}%
  \BibitemOpen
  \bibfield  {author} {\bibinfo {author} {\bibfnamefont {N.~B.}\ \bibnamefont
  {Budanur}}\ and\ \bibinfo {author} {\bibfnamefont {M.}~\bibnamefont
  {Fleury}},\ }\bibfield  {title} {\bibinfo {title} {State space geometry of
  the chaotic pilot-wave hydrodynamics},\ }\href
  {https://doi.org/10.1063/1.5058279} {\bibfield  {journal} {\bibinfo
  {journal} {Chaos: An Interdisciplinary Journal of Nonlinear Science}\
  }\textbf {\bibinfo {volume} {29}},\ \bibinfo {pages} {013122} (\bibinfo
  {year} {2019})},\ \Eprint
  {https://arxiv.org/abs/https://doi.org/10.1063/1.5058279}
  {https://doi.org/10.1063/1.5058279} \BibitemShut {NoStop}%
\bibitem [{\citenamefont {Osinga}\ and\ \citenamefont
  {Krauskopf}(2002)}]{Osinga2002}%
  \BibitemOpen
  \bibfield  {author} {\bibinfo {author} {\bibfnamefont {H.~M.}\ \bibnamefont
  {Osinga}}\ and\ \bibinfo {author} {\bibfnamefont {B.}~\bibnamefont
  {Krauskopf}},\ }\bibfield  {title} {\bibinfo {title} {Visualizing the
  structure of chaos in the {L}orenz system},\ }\href@noop {} {\bibfield
  {journal} {\bibinfo  {journal} {Comput. Graph.}\ }\textbf {\bibinfo {volume}
  {26}},\ \bibinfo {pages} {815} (\bibinfo {year} {2002})}\BibitemShut
  {NoStop}%
\bibitem [{\citenamefont {Osinga}(2018)}]{Osinga2018}%
  \BibitemOpen
  \bibfield  {author} {\bibinfo {author} {\bibfnamefont {H.~M.}\ \bibnamefont
  {Osinga}},\ }\bibfield  {title} {\bibinfo {title} {Understanding the geometry
  of dynamics: the stable manifold of the {L}orenz system},\ }\href@noop {}
  {\bibfield  {journal} {\bibinfo  {journal} {J. R. Soc. N. Z.}\ }\textbf
  {\bibinfo {volume} {48}},\ \bibinfo {pages} {203} (\bibinfo {year}
  {2018})}\BibitemShut {NoStop}%
\bibitem [{\citenamefont {Corless}(1992)}]{Gaussmap}%
  \BibitemOpen
  \bibfield  {author} {\bibinfo {author} {\bibfnamefont {R.~M.}\ \bibnamefont
  {Corless}},\ }\bibfield  {title} {\bibinfo {title} {Continued fractions and
  chaos},\ }\href@noop {} {\bibfield  {journal} {\bibinfo  {journal} {Am. Math.
  Mon.}\ }\textbf {\bibinfo {volume} {99}},\ \bibinfo {pages} {203} (\bibinfo
  {year} {1992})}\BibitemShut {NoStop}%
\bibitem [{\citenamefont {Aizawa}(1982)}]{Aizawa1982}%
  \BibitemOpen
  \bibfield  {author} {\bibinfo {author} {\bibfnamefont {Y.}~\bibnamefont
  {Aizawa}},\ }\bibfield  {title} {\bibinfo {title} {Global aspects of the
  dissipative dynamical systems. {I}: Statistical identification and fractal
  properties of the {L}orenz chaos},\ }\href@noop {} {\bibfield  {journal}
  {\bibinfo  {journal} {Prog. Theor. Phys.}\ }\textbf {\bibinfo {volume}
  {68}},\ \bibinfo {pages} {64} (\bibinfo {year} {1982})}\BibitemShut {NoStop}%
\bibitem [{\citenamefont {Aicardi}\ and\ \citenamefont
  {Borsellino}(1987)}]{Aicardi1987}%
  \BibitemOpen
  \bibfield  {author} {\bibinfo {author} {\bibfnamefont {F.}~\bibnamefont
  {Aicardi}}\ and\ \bibinfo {author} {\bibfnamefont {A.}~\bibnamefont
  {Borsellino}},\ }\bibfield  {title} {\bibinfo {title} {Statistical properties
  of flip-flop processes associated to the chaotic behavior of systems with
  strange attractors},\ }\href@noop {} {\bibfield  {journal} {\bibinfo
  {journal} {Biol. Cybern.}\ }\textbf {\bibinfo {volume} {55}},\ \bibinfo
  {pages} {377} (\bibinfo {year} {1987})}\BibitemShut {NoStop}%
\bibitem [{\citenamefont {Sancho}(1984)}]{Sancho1984}%
  \BibitemOpen
  \bibfield  {author} {\bibinfo {author} {\bibfnamefont {J.~M.}\ \bibnamefont
  {Sancho}},\ }\bibfield  {title} {\bibinfo {title} {Stochastic processes
  driven by dichotomous {M}arkov noise: Some exact dynamical results},\
  }\href@noop {} {\bibfield  {journal} {\bibinfo  {journal} {J. Math. Phys.}\
  }\textbf {\bibinfo {volume} {25}},\ \bibinfo {pages} {354} (\bibinfo {year}
  {1984})}\BibitemShut {NoStop}%
\bibitem [{\citenamefont {Brown}(1828)}]{Brown1828}%
  \BibitemOpen
  \bibfield  {author} {\bibinfo {author} {\bibfnamefont {R.}~\bibnamefont
  {Brown}},\ }\bibfield  {title} {\bibinfo {title} {{XXVII.} {A} brief account
  of microscopical observations made in the months of {J}une, {J}uly and
  {A}ugust 1827, on the particles contained in the pollen of plants; and on the
  general existence of active molecules in organic and inorganic bodies},\
  }\href {https://doi.org/10.1080/14786442808674769} {\bibfield  {journal}
  {\bibinfo  {journal} {Philos. Mag.}\ }\textbf {\bibinfo {volume} {4}},\
  \bibinfo {pages} {161} (\bibinfo {year} {1828})}\BibitemShut {NoStop}%
\bibitem [{\citenamefont {Brown}(1829)}]{Brown1829}%
  \BibitemOpen
  \bibfield  {author} {\bibinfo {author} {\bibfnamefont {R.}~\bibnamefont
  {Brown}},\ }\bibfield  {title} {\bibinfo {title} {{XXIV.} {A}dditional
  remarks on active molecules},\ }\href
  {https://doi.org/10.1080/14786442908675115} {\bibfield  {journal} {\bibinfo
  {journal} {Philos. Mag.}\ }\textbf {\bibinfo {volume} {6}},\ \bibinfo {pages}
  {161} (\bibinfo {year} {1829})}\BibitemShut {NoStop}%
\bibitem [{\citenamefont {Beck}(1996)}]{BECK1996419}%
  \BibitemOpen
  \bibfield  {author} {\bibinfo {author} {\bibfnamefont {C.}~\bibnamefont
  {Beck}},\ }\bibfield  {title} {\bibinfo {title} {Dynamical systems of
  {L}angevin type},\ }\href@noop {} {\bibfield  {journal} {\bibinfo  {journal}
  {Physica A}\ }\textbf {\bibinfo {volume} {233}},\ \bibinfo {pages} {419 }
  (\bibinfo {year} {1996})}\BibitemShut {NoStop}%
\bibitem [{\citenamefont {Shimizu}(1993)}]{SHIMIZU1993113}%
  \BibitemOpen
  \bibfield  {author} {\bibinfo {author} {\bibfnamefont {T.}~\bibnamefont
  {Shimizu}},\ }\bibfield  {title} {\bibinfo {title} {Chaotic force in
  {B}rownian motion},\ }\href@noop {} {\bibfield  {journal} {\bibinfo
  {journal} {Physica A}\ }\textbf {\bibinfo {volume} {195}},\ \bibinfo {pages}
  {113 } (\bibinfo {year} {1993})}\BibitemShut {NoStop}%
\bibitem [{\citenamefont {Chew}\ and\ \citenamefont
  {Ting}(2002)}]{CHEW2002275}%
  \BibitemOpen
  \bibfield  {author} {\bibinfo {author} {\bibfnamefont {L.}~\bibnamefont
  {Chew}}\ and\ \bibinfo {author} {\bibfnamefont {C.}~\bibnamefont {Ting}},\
  }\bibfield  {title} {\bibinfo {title} {Microscopic chaos and {G}aussian
  diffusion processes},\ }\href@noop {} {\bibfield  {journal} {\bibinfo
  {journal} {Physica A}\ }\textbf {\bibinfo {volume} {307}},\ \bibinfo {pages}
  {275 } (\bibinfo {year} {2002})}\BibitemShut {NoStop}%
\bibitem [{\citenamefont {Tref\'an}\ \emph {et~al.}(1992)\citenamefont
  {Tref\'an}, \citenamefont {Grigolini},\ and\ \citenamefont
  {West}}]{Trefan1992}%
  \BibitemOpen
  \bibfield  {author} {\bibinfo {author} {\bibfnamefont {G.}~\bibnamefont
  {Tref\'an}}, \bibinfo {author} {\bibfnamefont {P.}~\bibnamefont
  {Grigolini}},\ and\ \bibinfo {author} {\bibfnamefont {B.~J.}\ \bibnamefont
  {West}},\ }\bibfield  {title} {\bibinfo {title} {Deterministic {B}rownian
  motion},\ }\href@noop {} {\bibfield  {journal} {\bibinfo  {journal} {Phys.
  Rev. A}\ }\textbf {\bibinfo {volume} {45}},\ \bibinfo {pages} {1249}
  (\bibinfo {year} {1992})}\BibitemShut {NoStop}%
\bibitem [{\citenamefont {Huerta-Cuellar}\ \emph {et~al.}(2014)\citenamefont
  {Huerta-Cuellar}, \citenamefont {Jiménez-López}, \citenamefont
  {Campos-Cantón},\ and\ \citenamefont {Pisarchik}}]{HUERTACUELLAR20142740}%
  \BibitemOpen
  \bibfield  {author} {\bibinfo {author} {\bibfnamefont {G.}~\bibnamefont
  {Huerta-Cuellar}}, \bibinfo {author} {\bibfnamefont {E.}~\bibnamefont
  {Jiménez-López}}, \bibinfo {author} {\bibfnamefont {E.}~\bibnamefont
  {Campos-Cantón}},\ and\ \bibinfo {author} {\bibfnamefont {A.}~\bibnamefont
  {Pisarchik}},\ }\bibfield  {title} {\bibinfo {title} {An approach to generate
  deterministic {B}rownian motion},\ }\href@noop {} {\bibfield  {journal}
  {\bibinfo  {journal} {Commun. Nonlinear Sci. Numer. Simul.}\ }\textbf
  {\bibinfo {volume} {19}},\ \bibinfo {pages} {2740 } (\bibinfo {year}
  {2014})}\BibitemShut {NoStop}%
\bibitem [{\citenamefont {Lei}\ and\ \citenamefont {Mackey}(2011)}]{Lei2011}%
  \BibitemOpen
  \bibfield  {author} {\bibinfo {author} {\bibfnamefont {J.}~\bibnamefont
  {Lei}}\ and\ \bibinfo {author} {\bibfnamefont {M.~C.}\ \bibnamefont
  {Mackey}},\ }\bibfield  {title} {\bibinfo {title} {Deterministic {B}rownian
  motion generated from differential delay equations},\ }\href@noop {}
  {\bibfield  {journal} {\bibinfo  {journal} {Phys. Rev. E}\ }\textbf {\bibinfo
  {volume} {84}},\ \bibinfo {pages} {041105} (\bibinfo {year}
  {2011})}\BibitemShut {NoStop}%
\bibitem [{\citenamefont {Mol{\'a}{\v{c}}ek}(2013)}]{phdthesismolacek}%
  \BibitemOpen
  \bibfield  {author} {\bibinfo {author} {\bibfnamefont {J.}~\bibnamefont
  {Mol{\'a}{\v{c}}ek}},\ }\emph {\bibinfo {title} {Bouncing and walking
  droplets : towards a hydrodynamic pilot-wave theory}},\ \href@noop {} {Ph.D.
  thesis},\ \bibinfo  {school} {Massachusetts Institute of Technology}
  (\bibinfo {year} {2013})\BibitemShut {NoStop}%
\end{thebibliography}%

\end{document}